\documentclass[twoside]{article}
\usepackage[preprint]{aistats2027}
\usepackage[round]{natbib}

\usepackage{amsmath,amssymb,amsfonts,amsthm,mathtools}
\usepackage{booktabs,multirow,array}
\usepackage{enumitem}
\usepackage{graphicx}
\usepackage{xcolor}
\usepackage{hyperref}
\usepackage{microtype}
\usepackage{algorithm}
\usepackage{algcompatible}
\usepackage{caption}
\usepackage{subcaption}
\usepackage{float}
\usepackage{xspace}
\hypersetup{colorlinks=true,linkcolor=blue,citecolor=blue,urlcolor=blue}

\newtheorem{assumption}{Assumption}

\newtheorem{proposition}{Proposition}
\newtheorem{lemma}{Lemma}
\newtheorem{theorem}{Theorem}
\newtheorem{corollary}{Corollary}
\newtheorem{remark}{Remark}

\newcommand{\ket}[1]{\left|#1\right\rangle}
\newcommand{\bra}[1]{\left\langle#1\right|}
\newcommand{\braket}[2]{\left\langle #1\middle|#2\right\rangle}

\newcommand{\DAQNM}{\textsc{DA-QNM}\xspace}

\newcommand{\DAMuon}{\textsc{DA-Muon}\xspace}

\usepackage{amsmath,amsfonts,bm}

\def\eqref#1{equation~\ref{#1}}

\def\1{\bm{1}}

\DeclareMathAlphabet{\mathsfit}{\encodingdefault}{\sfdefault}{m}{sl}
\SetMathAlphabet{\mathsfit}{bold}{\encodingdefault}{\sfdefault}{bx}{n}

\DeclareMathOperator*{\argmax}{arg\,max}
\DeclareMathOperator*{\argmin}{arg\,min}

\begin{document}

\twocolumn[



 \aistatstitle{Quantum Natural Muon: Spectral-Natural Optimization for Variational Quantum Circuits}
    
    \aistatsauthor{Duong The Do \And Duong Tung Nguyen}
    
    \aistatsaddress{School of ECEE \\Arizona State University \\Tempe, AZ 85281, USA \And  School of ECEE \\ Arizona State University \\ Tempe, AZ 85281, USA} 
]

\begin{abstract}
Training parameterized quantum circuits requires optimization methods that account for two distinct structures: the information geometry of quantum states and the matrix or tensor structure induced by circuit parameterization. We introduce \emph{Quantum Natural Muon} (QNM), a spectral natural-gradient framework that applies polar normalization to circuit-semantic parameter blocks after damped quantum Fisher preconditioning. This ordering is essential: quantum Fisher preconditioning and blockwise polar normalization do not commute in general, so QNM is not an order-free composition of quantum natural gradient (QNG) and Muon. We show that Frobenius-matched polar mixing preserves active singular subspaces while raising a spectral floor and improving active conditioning, and that the resulting direction remains a controlled perturbation of QNG. For its distance-aware realization (DA-QNM), we further establish alignment, descent, and nonconvex stationarity guarantees for the normalized update, together with perturbation bounds under gradient and QFIM estimation error. Experiments on variational quantum eigensolving, weighted QAOA-MaxCut, and supervised quantum classification isolate the roles of quantum geometry, circuit-semantic spectral structure, and radius control. The strongest gains arise when the spectral blocks match the circuit structure, supporting the central principle that quantum-state and structured spectral geometries provide complementary information for variational quantum optimization.
\end{abstract} 
\section{Introduction}
\label{sec:intro}
\vspace{-0.3cm}
\noindent
Parameterized quantum circuits (PQCs) underlie variational quantum
eigensolving, quantum approximate optimization, and quantum learning
\citep{cerezo2021variational,benedetti2019parameterized,
peruzzo2014variational,farhi2014quantum}. Their optimization has two
distinct structures. First, Euclidean parameter distance does not generally
reflect distance between the induced quantum states: different parameter
directions can have very different local state sensitivities
\citep{amari1998natural,stokes2020quantum}. Second, PQC parameters are not
unstructured coordinates; circuit architectures organize them by layers,
qubits, gate families, Hamiltonian terms, graph edges, or data-reuploading
blocks
\citep{benedetti2019parameterized,cerezo2021variational,
farhi2014quantum,perezsalinas2020data}. Variational quantum optimization
therefore couples \emph{quantum-state information geometry} with
\emph{circuit-induced parameter structure}.

\noindent
Existing optimizers address these structures separately. Adam and related
Euclidean methods adapt coordinate-wise scales but do not account for
quantum-state geometry \citep{duchi2011adaptive,kingma2015adam}. Quantum
natural gradient (QNG) instead preconditions gradients with the quantum
Fisher information matrix (QFIM), calibrating updates to local
state sensitivity \citep{stokes2020quantum}; however, the resulting
natural-gradient vector does not control how update magnitude is distributed
across circuit-structured matrix or tensor blocks. Muon addresses a
different geometry by reshaping momentum into matrices and applying polar
normalization to their singular spectra
\citep{jordan2024muon,bernstein2025modular}. Distance-Aware Muon further
separates normalized direction selection from scalar displacement control
\citep{demidovich2026distanceaware}, but neither method incorporates
quantum-state geometry.

\noindent
Exploiting both geometries creates several nontrivial design problems.
Spectral normalization can be applied before or after QFIM preconditioning,
but these operations do not commute in general and therefore define
different update directions. Full polar normalization also equalizes active
singular values at the cost of discarding block-scale information, whereas
an aggressive spectral modification can move the update far enough from
QNG to lose its descent property. Finally, normalizing the resulting
direction separates direction selection from displacement magnitude, so the
step radius must be controlled independently. The central question is
therefore not how to concatenate QNG and Muon, but \emph{how to exploit
circuit-level spectral structure while preserving the quantum-geometric and
descent information of the natural-gradient direction}.

\noindent
We introduce \emph{Quantum Natural Muon} (QNM), a spectral
natural-gradient framework for structured variational quantum optimization.
Given gradient $g_t$, QNM first computes the damped natural-gradient vector
$u_t=(\mathbf F_t+\lambda\mathbf I)^{-1}g_t$ and accumulates momentum in
this quantum-geometric space. It then reshapes the momentum according to
circuit semantics and mixes each block with a Frobenius-matched polar
factor, balancing its active spectrum without replacing its block-scale
information. This ordering is deliberate: because QFIM preconditioning and
blockwise polar normalization do not commute, applying spectral
normalization before state-sensitivity calibration generally defines a
different optimizer. QNM instead performs spectral direction selection
\emph{after} quantum-geometric calibration. Its distance-aware realization,
DA-QNM, normalizes the resulting spectral-natural direction and controls
the final displacement with an adaptive radius; the radius rule itself is
not claimed as new.

\noindent
Our analysis characterizes both what this transformation changes and what
it preserves. The polar factor solves a spectral-norm-constrained local
direction-selection problem. With Frobenius-matched scaling, polar
mixing preserves the active singular subspaces of a structured
natural-gradient block while raising a spectral floor and improving its
active spectral conditioning. We then bound the deviation of the resulting
direction from QNG and derive sufficient conditions for gradient alignment
and descent. For the normalized DA-QNM update, we connect the radius
generated by the algorithm directly to descent and establish an
$O(1/T)$ nonconvex stationarity guarantee. We additionally bound the
sensitivity of the natural-gradient direction to gradient and QFIM
estimation errors, quantifying the stabilizing role of damping.

\noindent
We evaluate QNM on VQE, weighted generalized QAOA-MaxCut, and supervised
quantum classification, which expose layer-, gate/graph-, and
data-reuploading structure, respectively
\citep{peruzzo2014variational,cerezo2021variational,farhi2014quantum,
havlicek2019supervised,perezsalinas2020data}. Rather than only comparing
final objectives, our ablations isolate quantum preconditioning, spectral
normalization, semantic block choice, and radius control. On the weighted
$n=10$, $p=3$ QAOA instance, QNG reduces the Adam optimality gap by
$43.0\%$, while gate-structured QNM reduces it by $96.9\%$ and DA-QNM by
$97.0\%$. In contrast, Euclidean, layer-wise, and global spectral variants
do not obtain the same gain, showing that spectral normalization alone is
insufficient. Consistent with the predicted mechanism, polar mixing reduces
the measured active condition number of a representative structured update
from $10.13$ to $1.85$ while increasing its effective rank from $2.22$ to
$2.84$. The gains are therefore strongest when quantum-state geometry and
circuit-semantic spectral structure are jointly informative.

\noindent
Our main contributions are:
\begin{enumerate}[itemsep=2pt,parsep=0pt,topsep=1pt]
    \item \textbf{Structured spectral natural gradients.}
    We introduce QNM, which performs circuit-semantic spectral direction
    selection after damped QFIM preconditioning. Because QFIM
    preconditioning and polar normalization do not commute, their ordering
    defines different optimizers; QNM deliberately regularizes the
    structured spectrum only after state-sensitivity calibration.

    \item \textbf{Spectral and optimization guarantees.}
    We show that Frobenius-matched polar mixing preserves active singular
    subspaces, raises a spectral floor, and improves active spectral
    conditioning while remaining a controlled perturbation of QNG, yielding
    explicit alignment and descent conditions.

    \item \textbf{Normalized-update convergence and estimation sensitivity.}
    For DA-QNM, we connect the algorithm's generated radius to descent and
    $O(1/T)$ nonconvex stationarity and derive perturbation bounds under
    gradient and QFIM estimation error.

    \item \textbf{Mechanism-driven empirical validation.}
    Across VQE, weighted QAOA, and supervised quantum classification,
    component and block-structure ablations show that the gains arise from
    jointly exploiting quantum-state geometry and circuit-semantic spectral
    structure rather than generic normalization or radius control.
\end{enumerate}

\vspace{-0.3cm}
\section{Preliminaries}
\label{sec:background}
\vspace{-0.35cm}

\noindent
We formalize the two structures used by QNM: quantum-state geometry,
represented by the quantum Fisher information matrix (QFIM), and
circuit-semantic block structure.

\vspace{-0.25cm}
\subsection{Quantum Models and Circuit Structure}
\label{subsec:pqms}

\noindent
Let $\mathbf U(\theta)$ be a parameterized quantum circuit with
$\theta\in\mathbb R^d$ and
$\ket{\psi(\theta)}=\mathbf U(\theta)\ket{0}^{\otimes n}$.
We consider Hamiltonian objectives
$C(\theta)=\bra{\psi(\theta)}\mathbf H\ket{\psi(\theta)}$, covering VQE
and QAOA-type problems
\citep{peruzzo2014variational,cerezo2021variational,farhi2014quantum,
hadfield2019from}, as well as empirical losses for supervised variational
quantum models
\citep{schuld2019quantum,havlicek2019supervised,
perezsalinas2020data}.

\noindent
Across these settings, quantum parameters have circuit-induced structure:
hardware-efficient ansatzes organize parameters by layer, qubit, and
rotation axis; generalized QAOA by depth, graph edge, and node; and
data-reuploading circuits by layer and encoding block. Let
$\mathcal S=\{1,\ldots,S\}$ index parameter blocks. For each
$s\in\mathcal S$, let $\mathcal I_s\subseteq\{1,\ldots,d\}$ index a block,
$\theta^{(s)}=\theta_{\mathcal I_s}\in\mathbb R^{d_s}$, and define a
circuit-semantic reshape
\begin{align}
    \mathcal R_s:\mathbb R^{d_s}\rightarrow\mathbb R^{a_s\times b_s},
    \quad
    \mathbf M^{(s)}=\mathcal R_s(\theta^{(s)}).
\end{align}
The maps $\mathcal R_s$ determine which circuit-level correlations are
exposed to spectral normalization; task-specific choices are given in
Section~\ref{sec:experiments} and Appendix~\ref{app:experiment_details}. 

\subsection{Quantum Natural Gradient}
\label{subsec:qng}
\vspace{-0.2cm}
\noindent
For pure states, the QFIM is the real part of the quantum geometric tensor
\citep{provost1980riemannian,stokes2020quantum},
\begin{align}
[\mathbf F(\theta)]_{ij}
=\operatorname{Re}\!\left[
\braket{\partial_i\psi}{\partial_j\psi}
-\braket{\partial_i\psi}{\psi}
 \braket{\psi}{\partial_j\psi}
\right].
\label{eq:qfim}
\end{align}
It measures local quantum-state sensitivity to parameter perturbations.
With $g_t=\nabla C(\theta_t)$ and damping $\lambda>0$, define $\mathbf G_t=\mathbf F_t+\lambda\mathbf I$ and $u_t=\mathbf G_t^{-1}g_t .$
Here $u_t$ is the damped QNG vector and $-u_t$ is the corresponding
descent direction \citep{stokes2020quantum}. Additional geometric details
and the equivalent local variational characterization are given in
Appendix~\ref{app:qng}.

\noindent
QNG calibrates parameter motion to quantum-state geometry, but imposes no
spectral constraint on the circuit-semantic matrices
$\mathcal R_s(u_t^{(s)})$. QFIM preconditioning therefore need not remove
spectral anisotropy within the resulting structured update blocks. This is
an update-structure property, distinct from conditioning of the objective
landscape.

\subsection{Polar Spectral Normalization}
\label{subsec:muon}
\vspace{-0.2cm}
\noindent
Muon-style methods act instead on matrix-update geometry
\citep{jordan2024muon,bernstein2025modular}. For a nonzero block
$\mathbf M=\mathbf U\boldsymbol{\Sigma}\mathbf V^\top$ with compact SVD,
define $\operatorname{Polar}(\mathbf M)=\mathbf U\mathbf V^\top.$
The polar factor preserves the active left and right singular subspaces
while setting all active singular values to one. It therefore removes
active spectral imbalance without rotating those subspaces. Its
spectral-norm variational characterization, used in
Section~\ref{sec:theory} to interpret QNM as constrained matrix direction
selection, is detailed in Appendix~\ref{app:muon}.

\noindent
Polar normalization thus addresses a geometry different from QNG: QNG
calibrates directions by quantum-state sensitivity, whereas the polar
operation controls the active spectrum of a structured matrix update.
Full polar normalization, however, replaces all active singular values by
one and therefore discards their relative magnitudes and the original
block scale. This motivates the scale-preserving polar mixing introduced
next. Moreover, because QFIM preconditioning and blockwise polar
normalization generally do not commute, their ordering is consequential.
Section~\ref{sec:method} constructs QNM by applying spectral normalization
\emph{after} quantum preconditioning.

\section{Quantum Natural Muon} 
\label{sec:method}
\vspace{-0.2cm}
 \noindent
We now define Quantum Natural Muon (QNM) and its distance-aware realization
(DA-QNM). QNM applies spectral normalization to momentum accumulated in
natural-gradient space; DA-QNM additionally normalizes the resulting
spectral-natural direction and controls its displacement through a
distance-aware radius. The update pipeline is
    \begin{align}
    \label{eq:daqnm_pipeline}
        g_t &\to u_t=(\mathbf F_t+\lambda\mathbf I)^{-1}g_t \to m_t \to \{\mathbf M_t^{(s)}\}_{s\in\mathcal S} \\ &\to \{\mathbf D_t^{(s)}\}_{s\in\mathcal S} \to d_t \to q_t \to \theta_{t+1}=\theta_t-r_t q_t . \nonumber
    \end{align}
    Here $g_t=\nabla C(\theta_t)$ is the Euclidean gradient, $u_t$ is the damped natural-gradient vector, $m_t$ is natural-gradient momentum, $\mathbf M_t^{(s)}$ are circuit-semantic matrix blocks, $\mathbf D_t^{(s)}$ are polar-mixed blocks, $d_t$ is the concatenated direction, $q_t$ is the normalized direction, and $r_t$ is the distance-aware radius.

\subsection{Natural-Gradient Momentum Blocks}
\label{subsec:natural_gradient_momentum_block_structure}
\vspace{-0.2cm}
    Rather than forming $\mathbf G_t^{-1}$ explicitly, \DAQNM computes the damped natural-gradient vector by solving $\mathbf G_t u_t \!=\! g_t$. The final parameter update moves in the negative direction of the transformed vector, so $u_t$ is the natural-gradient vector and $-u_t$ is the corresponding descent displacement. \DAQNM accumulates momentum in natural-gradient space:
    \begin{align}
    \label{eq:natural_momentum}
        m_t = \beta m_{t-1} + (1-\beta)u_t, \quad \beta\in[0,1).
    \end{align}
    Thus, the subsequent spectral normalization acts on a quantum-geometrically calibrated direction rather than on the raw Euclidean gradient.

    For each block $s\in\mathcal S$, let $m_t^{(s)}\in\mathbb R^{d_s}$ denote the corresponding subvector of $m_t$. Using the circuit-semantic reshape map $\mathcal R_s$, \DAQNM forms the matrix block
    \begin{align}
    \label{eq:block_reshape}
        \mathbf M_t^{(s)} = \mathcal R_s(m_t^{(s)}).
    \end{align}
    The choice of $\mathcal R_s$ determines which circuit-induced correlations are spectrally regularized. In our experiments, we use layer-wise blocks for VQE, gate-wise cost/mixer blocks for generalized QAOA, and data-reuploading blocks for supervised quantum classifiers.

\subsection{Blockwise Spectral Normalization}
\label{subsec:blockwise_spectral_normalization}
\vspace{-0.2cm}
    For each structured momentum block $\mathbf M_t^{(s)}$, let
    \begin{align}
    \label{eq:block_svd}
        \mathbf M_t^{(s)} = \mathbf U_t^{(s)} \boldsymbol{\Sigma}_t^{(s)} \mathbf V_t^{(s)\top}
    \end{align}
    be its compact singular value decomposition. If $\mathbf M_t^{(s)}\neq 0$, define its polar factor as
    \begin{align}
    \label{eq:block_polar}
        \mathbf P_t^{(s)} = \operatorname{Polar}(\mathbf M_t^{(s)}) = \mathbf U_t^{(s)}\mathbf V_t^{(s)\top}.
    \end{align}
    The polar factor preserves the active left and right singular subspaces of $\mathbf M_t^{(s)}$, while setting its nonzero singular values to one. Replacing $\mathbf M_t^{(s)}$ entirely by its polar factor can be too aggressive because it discards block magnitude information. \DAQNM therefore uses Frobenius-preserving polar mixing:
    \begin{align}
    \label{eq:block_mix_revised}
        \mathbf D_t^{(s)} = (1-\alpha)\mathbf M_t^{(s)} + \alpha c_t^{(s)}\mathbf P_t^{(s)}, \quad \alpha\in[0,1].
    \end{align}
    The scalar $c_t^{(s)}$ controls the scale of the polar component. With Frobenius-norm preservation, we set
    \begin{align}
    \label{eq:frobenius_scale}
        c_t^{(s)} = \frac{\|\mathbf M_t^{(s)}\|_F} {\|\mathbf P_t^{(s)}\|_F} = \frac{\|\mathbf M_t^{(s)}\|_F} {\sqrt{\operatorname{rank}(\mathbf M_t^{(s)})}} .
    \end{align}
    If $\mathbf M_t^{(s)} \!=\! 0$, we set $\mathbf D_t^{(s)} \!=\! 0$. The scaling in (\ref{eq:frobenius_scale}) matches the Frobenius norm of
the polar component to that of the original block while redistributing
magnitude across active singular directions.
    The blockwise matrices are then vectorized and concatenated $d_t \!=\! \operatorname{vec}\! \big(\{\mathbf D_t^{(s)}\}_{s \in \mathcal S} \big)$. This vector is the fixed-scale spectral-natural direction. The distance-aware step further normalizes $d_t$ and chooses a 
    scalar radius.

\subsection{Distance-Aware Radius Control}
\label{subsec:distance_aware_radius}
\vspace{-0.2cm}
    Polar mixing controls the structured direction but partially changes magnitude information. \DAQNM therefore separates direction from step length by defining the normalized spectral-natural direction
    \begin{align}
    \label{eq:normalized_qnm_direction}
        q_t=\frac{d_t}{\|d_t\|_2+\epsilon},\quad\epsilon>0,
    \end{align}
    and updating $\theta_{t+1} \!=\! \theta_t\!-\!r_t q_t$.
    Distance-aware radius is
    \begin{align}
    \label{eq:radius_revised}
        r_t \!=\! \min \!\Big\{\!\eta\|m_t\|_2, \Delta_0 \!+\! \rho \!\! \max_{0\leq k\leq t} \!\|\theta_k \!-\! \theta_0\|_2, R_{\max} \!\Big\}.\!
    \end{align}
    The first term ties the step length to the natural-gradient momentum magnitude. The second adapts the radius to the trajectory displacement from initialization. The third prevents extreme normalized steps.

    We do not claim that the radius rule itself is new; rather, \DAQNM uses distance-aware scaling to control the step length after QFIM-preconditioned spectral direction selection. Thus, the core contribution is the spectral-natural direction, while distance-aware control stabilizes the normalized displacement produced by that direction.
    This decomposition separates the three roles of \DAQNM: \emph{QFIM preconditioning} calibrates the direction to quantum-state geometry, \emph{polar mixing} balances the active spectrum in circuit-semantic blocks, and \emph{distance-aware radius} control determines the scalar displacement along the normalized direction.

\textbf{Why the order matters.}
    \DAQNM applies polar mixing after QFIM preconditioning. To formalize the ordering, consider a single block $s$ and define the blockwise metric-preconditioning operator
    \begin{align}
    \label{eq:block_metric_operator}
         \mathcal A_{t,s}(\mathbf M) = \mathcal R_s \left( \left[ \mathbf G_t^{-1} \iota_s \big( \operatorname{vec}(\mathbf M) \big) \right]^{(s)} \right),
    \end{align}
    where $\iota_s$ embeds a block vector into the full parameter space by zero-padding outside $s$, and $[\cdot]^{(s)}$ extracts block-$s$ coordinates. The operator $\mathcal A_{t,s}$ represents the effect of QFIM preconditioning on a matrix block after embedding it into the full parameter vector. In general,
    \begin{align}
    \label{eq:noncommuting_order}
        \operatorname{Polar} \big( \mathcal A_{t,s}(\mathbf M) \big) \neq \mathcal A_{t,s} \big( \operatorname{Polar}(\mathbf M) \big).
    \end{align}

    \begin{remark}
        QFIM preconditioning and blockwise polar normalization do not commute in general. \DAQNM first computes a quantum-geometrically calibrated direction and then regularizes its structured spectrum, producing a spectral natural-gradient update distinct from both QNG and Muon.
    \end{remark}
    
    \begin{algorithm}[t]
    \caption{Distance-Aware Quantum Natural Muon}
    \label{alg:daqnm}
    \begin{algorithmic}[1]
    \REQUIRE $\theta_0$, $\lambda$, $\beta$, $\alpha$, $\eta$, $\Delta_0$, $\rho$, $R_{\max}$, $\epsilon$, $\{\mathcal R_s\}_{s\in\mathcal S}$
    \STATE $m_{-1}\gets 0$
    \FOR{$t=0,1,2,\ldots$}
        \STATE Estimate gradient $g_t=\nabla C(\theta_t)$
        \STATE Compute QFIM $\mathbf F_t$, and set $\mathbf G_t=\mathbf F_t+\lambda\mathbf I$
        \STATE Solve $\mathbf G_tu_t=g_t$ for $u_t$
        \STATE $m_t\gets \beta m_{t-1}+(1-\beta)u_t$
        \FOR{each block $s\in\mathcal S$}
            \STATE $\mathbf M_t^{(s)}\gets \mathcal R_s(m_t^{(s)})$
            \IF{$\mathbf M_t^{(s)}=0$}
                \STATE $\mathbf D_t^{(s)}\gets 0$
            \ELSE
                \STATE $\mathbf P_t^{(s)}\gets \operatorname{Polar}(\mathbf M_t^{(s)})$
                \STATE $c_t^{(s)}\gets \|\mathbf M_t^{(s)}\|_F/\|\mathbf P_t^{(s)}\|_F$
                \STATE $\mathbf D_t^{(s)} \gets (1-\alpha)\mathbf M_t^{(s)} + \alpha c_t^{(s)}\mathbf P_t^{(s)}$
            \ENDIF
        \ENDFOR
        \STATE $d_t\gets \operatorname{vec}(\{\mathbf D_t^{(s)}\}_{s\in\mathcal S})$
        \STATE $q_t\gets d_t/(\|d_t\|_2+\epsilon)$
        \STATE $R_t^{\mathrm{dist}} \gets \Delta_0+\rho\max_{0\leq k\leq t}\|\theta_k-\theta_0\|_2$
        \STATE $r_t\gets \min\{\eta\|m_t\|_2,R_t^{\mathrm{dist}},R_{\max}\}$
        \STATE $\theta_{t+1}\gets \theta_t-r_tq_t$
    \ENDFOR
    \end{algorithmic}
    \end{algorithm}
\section{Theoretical Analysis}
\label{sec:theory}
\vspace{-0.25cm}
    This section formalizes why \DAQNM is a structured spectral natural-gradient method. The results establish four claims: polar factors solve local spectral-norm direction-selection problems; QFIM preconditioning and blockwise polar normalization do not commute in general; polar mixing improves active spectral conditioning; and the resulting update admits descent, stationarity, and estimation-robustness guarantees under standard assumptions. Complete proofs are deferred to Appendix~\ref{app:proofs}.

    \begin{assumption}[Smoothness]
    \label{assump:smooth}
        The objective $C$ is $L$-smooth, i.e., $C(\theta \!+\!\delta) \!\leq\! C(\theta) \!+\! \langle \nabla C(\theta), \delta\rangle \!+\! \frac{L}{2} \|\delta\|_2^2$ for all $\theta,\delta \!\in\! \mathbb R^d$.
    \end{assumption}

    \begin{assumption}[Damped QFIM boundedness]
    \label{assump:qfim}
        For all iterates, the damped QFIM $\mathbf G_t \!=\! \mathbf F_t \!+\! \lambda \mathbf I$, $\lambda \!>\! 0$ satisfies $\mu \mathbf I \!\preceq\! \mathbf G_t \!\preceq\! \Lambda \mathbf I$ for constants $0 \!<\! \mu \!\leq\! \Lambda \!<\! \infty$.
    \end{assumption}


    \begin{assumption}[Block decomposition]
    \label{assump:block_norm}
        The circuit-semantic blocks form a norm-consistent decomposition of the parameter coordinates. Specifically, for any $v \!\in\! \mathbb{R}^d$, let $v^{(s)} \!\in\! \mathbb{R}^{d_s}$ be its block-$s$ subvector and define $\mathbf M^{(s)} \!=\! \mathcal R_s(v^{(s)})$. We assume $\sum_{s\in\mathcal S} \|\mathbf M^{(s)}\|_F^2 \!=\! \|v\|_2^2$.
    \end{assumption}

    \begin{assumption}[Alignment of normalized direction]
    \label{assump:align}
        For the normalized \DAQNM direction $q_t$, there exists $\chi_t>0$ such that $\langle g_t,q_t\rangle \!\geq\! \chi_t\|g_t\|_2$.
    \end{assumption}

    \begin{assumption}[Lower bounded objective]
    \label{assump:lower_bound}
        The objective is bounded below: there exists $C_\star>-\infty$ such that $C(\theta) \!\geq\! C_\star$ for all $\theta\in\mathbb{R}^d$.
    \end{assumption}

    We first state the local direction-selection property of the polar factor.

    \begin{theorem}[Spectral-norm steepest direction]
    \label{thm:spectral_lmo}
        Let $\mathbf M\in\mathbb R^{a\times b}$ have compact SVD, $\mathbf M \!=\! \mathbf U\boldsymbol{\Sigma}\mathbf V^\top$ with rank $r$. If $\mathbf M \!\neq\! 0$, then its compact polar factor $\operatorname{Polar}(\mathbf M) \!=\! \mathbf U\mathbf V^\top$ solves $\operatorname{Polar}(\mathbf M) \in \argmax_{\|\mathbf D\|_2\leq1} \langle \mathbf M,\mathbf D\rangle_F$.
        The optimal value is $\|\mathbf M\|_\ast$. If $\mathbf M=0$, every feasible $\mathbf D$ is optimal and the optimal value is zero.
    \end{theorem}
    Theorem~\ref{thm:spectral_lmo} gives the polar step a local spectral trust-region interpretation: for a structured natural-gradient momentum $\mathbf M_t^{(s)}$, the polar factor is the steepest feasible matrix direction under a spectral-norm constraint. Thus, after QFIM preconditioning, QNM performs constrained direction
selection in each circuit-semantic block rather than merely normalizing
a matrix.

    We next formalize the non-commutativity claim introduced in Section~\ref{sec:method}. Recall the blockwise metric-preconditioning operator $\mathcal A_{t,s}$ from (\ref{eq:block_metric_operator}).
    
    \begin{proposition}[Non-commutativity of metric preconditioning and polar normalization]
    \label{prop:noncommute}
        In general,
        \begin{align}
        \label{eq:noncommute_statement}
            \operatorname{Polar} \big( \mathcal A_{t,s}(\mathbf M) \big) \neq \mathcal A_{t,s} \big( \operatorname{Polar}(\mathbf M) \big).
        \end{align}
        Thus, QFIM preconditioning and blockwise polar normalization do not commute in general. Equality would require restrictive compatibility between $\mathbf G_t^{-1}$ and the block reshape $\mathcal R_s$, such as isotropic action on the relevant block subspace.
    \end{proposition}

    Proposition~\ref{prop:noncommute} motivates the algorithmic design:
\DAQNM applies spectral normalization after quantum-metric calibration,
yielding a spectral natural-gradient direction distinct from QNG and Muon.
We next characterize how polar mixing modifies the active singular spectrum.
For a nonzero matrix $\mathbf M$ of rank $r$, define the active condition
number
$\kappa_{\mathrm{act}}(\mathbf M)
\!=\!\frac{\sigma_1(\mathbf M)}{\sigma_r(\mathbf M)}$,
    where $\sigma_1(\mathbf M)\geq\cdots\geq\sigma_r(\mathbf M)>0$ are 
    nonzero singular values.

    \begin{proposition}[Active spectrum of the polar factor]
    \label{prop:active}
        Let $\mathbf M\in\mathbb R^{a\times b}$. If $\mathbf M\neq0$, let $r=\operatorname{rank}(\mathbf M)$ and let $\mathbf P=\operatorname{Polar}(\mathbf M)$. Then all active singular values of $\mathbf P$ equal one. Consequently, $\kappa_{\mathrm{act}}(\mathbf P)=1$.
    \end{proposition}

    \begin{proposition}[Frobenius-matched spectral balancing]
    \label{prop:fro}
        Let $\mathbf M\in\mathbb{R}^{a\times b}$ be nonzero with rank $r$, let $\mathbf P \!=\! \operatorname{Polar}(\mathbf M)$ and define $c \!=\! \frac{\|\mathbf M\|_F}{\|\mathbf P\|_F}$. Then $\|c\mathbf P\|_F \!=\! \|\mathbf M\|_F$. Moreover, every active singular value of $c\mathbf P$ equals $\frac{\|\mathbf M\|_F}{\sqrt r}$. If $\mathbf M\!=\!0$, we use the convention $c\mathbf P\!=\!0$.
    \end{proposition}

Proposition~\ref{prop:fro} explains the Frobenius-matched scaling
used in \DAQNM. The scaling matches the polar component's Frobenius norm
to that of the original block while redistributing magnitude uniformly
across active singular directions.
    We next analyze the mixed block direction $\mathbf D$. The following result shows that polar mixing raises a floor on the active singular spectrum while preserving the active singular subspaces.

    \begin{proposition}[Spectral floor from polar mixing]
    \label{prop:floor}
        Let $\mathbf M\in\mathbb R^{a\times b}$ be nonzero with compact SVD, i.e., $\mathbf M = \mathbf U\boldsymbol{\Sigma}\mathbf V^\top$, $\boldsymbol{\Sigma} = \operatorname{diag}(\sigma_1,\ldots,\sigma_r)$ and $\sigma_i>0$. Let $\mathbf P \!=\! \operatorname{Polar}(\mathbf M) \!=\! \mathbf U\mathbf V^\top$ and define $\mathbf D \!=\! (1-\alpha)\mathbf M+\alpha c\mathbf P$. Then $\mathbf D$ has the same active left and right singular subspaces as $\mathbf M$, and its active singular values are $\widetilde{\sigma}_i \!=\! (1\!-\!\alpha)\sigma_i \!+\!\alpha c$.
        In particular, $\widetilde{\sigma}_i \geq \alpha c$ for all $i=1,\ldots,r$.
    \end{proposition}

    \begin{corollary}[Active condition-number improvement]
    \label{cor:cond}
        Suppose $\sigma_1\geq\cdots\geq\sigma_r>0$. Then,
        \begin{align}
        \label{eq:condition_improvement}
            \kappa_{\mathrm{act}}(\mathbf D) = \frac{(1-\alpha)\sigma_1+\alpha c} {(1-\alpha)\sigma_r+\alpha c} \leq \frac{\sigma_1}{\sigma_r} = \kappa_{\mathrm{act}}(\mathbf M).
        \end{align}
        The inequality is strict if $\alpha\!>\!0$, $c\!>\!0$, and $\sigma_1\!>\!\sigma_r$.
    \end{corollary}

    Proposition~\ref{prop:floor} and Corollary~\ref{cor:cond} show that polar mixing lifts weak active singular directions while preserving active singular subspaces. This is the spectral-conditioning mechanism underlying QNM.

    We now show that the fixed-scale spectral-natural direction remains close to QNG when the polar mixing coefficient is controlled. This is important because it ensures that spectral normalization does not arbitrarily destroy the natural-gradient direction.

    \begin{lemma}[Perturbation size of polar mixing]
    \label{lemma:pert}
        Let $\mathbf M\in\mathbb{R}^{a\times b}$ be nonzero, let $\mathbf P \!=\! \operatorname{Polar}(\mathbf M)$, $c \!=\! \frac{\|\mathbf M\|_F}{\|\mathbf P\|_F}$ and define $\mathbf D \!=\! (1\!-\!\alpha) \mathbf M \!+\!\alpha c \mathbf P$, $\alpha \!\in\! [0,1]$. Then, $\|\mathbf D \!-\! \mathbf M\|_F \!\leq\! 2\alpha\|\mathbf M\|_F$.
        If $\mathbf M=0$, we use $\mathbf D=0$, and the bound holds trivially.
    \end{lemma}

    \begin{theorem}[Descent for controlled spectral-natural perturbations]
    \label{thm:qnm_descent}
        Suppose Assumptions~\ref{assump:smooth} and~\ref{assump:qfim} hold, and suppose $g_t\neq 0$. Let $g_t \!=\! \nabla C(\theta_t)$, $u_t \!=\! \mathbf G_t^{-1}g_t$ be the damped natural-gradient vector. Suppose the fixed-scale spectral-natural direction $d_t$ satisfies $\|d_t-u_t\|_2 \!\leq\! \varepsilon_t\|u_t\|_2$. If $\varepsilon_t \!<\! \frac{\mu}{\Lambda}$ then $d_t$ is a descent vector, i.e., $\langle g_t,d_t\rangle>0$. Moreover, any step size satisfying $0 \!<\! \gamma \!<\! \frac{2\langle g_t,d_t\rangle} {L\|d_t\|_2^2}$ yields $C(\theta_t-\gamma d_t) < C(\theta_t)$.
    \end{theorem}
    
    Theorem~\ref{thm:qnm_descent} states that if the spectral-natural direction is a sufficiently small relative perturbation of QNG, then it preserves descent. The threshold depends on the damped QFIM condition ratio $\mu/\Lambda$.

    \begin{corollary}[Sufficient polar-mixing condition]
    \label{cor:polar_descent}
        Consider the no-momentum single-block case, and suppose the reshape map is norm-consistent as in Assumption~\ref{assump:block_norm}. Let $\mathbf M$ be the reshape of $u_t$, and $\mathbf D$ the corresponding Frobenius-preserving polar-mixed block with mixing coefficient $\alpha$. If $2\alpha\!<\!\frac{\mu}{\Lambda}$ then the fixed-scale spectral-natural direction $d_t$ is guaranteed to be a descent vector.
    \end{corollary}

    \begin{corollary}[Descent with natural-gradient momentum]
    \label{cor:momentum_descent}
        Suppose Assumptions~\ref{assump:smooth}--\ref{assump:block_norm} hold. Let $u_t \!=\! \mathbf G_t^{-1}g_t$ be the damped natural-gradient vector, and suppose the natural-gradient momentum satisfies $\|m_t \!-\! u_t\|_2 \!\leq\! \xi_t\|u_t\|_2$ for some $\xi_t\geq0$. Let $d_t$ be obtained by applying Frobenius-preserving polar mixing with coefficient $\alpha$ to the circuit-semantic block decomposition of $m_t$. Then $\|d_t \!-\! u_t\|_2 \!\leq\! \left(2\alpha(1 \!+\! \xi_t) \!+\! \xi_t \right) \|u_t\|_2$.
        If $2\alpha(1 \!+\! \xi_t) \!+\! \xi_t \!<\! \frac{\mu}{\Lambda}$ then $d_t$ is a descent vector, i.e., $\langle g_t,d_t\rangle \!>\!0$.
    \end{corollary}
    
    Corollary~\ref{cor:momentum_descent} connects the analysis to the actual \DAQNM update, where polar mixing is applied to natural-gradient momentum $m_t$. When $\xi_t=0$, the result reduces to the no-momentum condition $2\alpha \!<\! \mu/\Lambda$. When $\xi_t>0$, descent is guaranteed provided the momentum remains close enough to the current damped natural-gradient direction.

    \begin{corollary}[Alignment of the normalized DA-QNM direction]
    \label{cor:normalized_alignment}
        Suppose Assumption~\ref{assump:qfim} holds and $g_t \!\neq\! 0$. Let $u_t \!=\! \mathbf G_t^{-1}g_t$, and suppose the fixed-scale spectral-natural direction satisfies $\|d_t \!-\! u_t\|_2 \!\le\! \zeta_t \|u_t\|_2$ for some $\zeta_t \!<\! \mu/\Lambda$. Let $q_t \!=\! \frac{d_t}{\|d_t\|_2 \!+\! \epsilon}$ where $\epsilon \!\le\! \tau_t\|u_t\|_2$ for some $\tau_t \!\ge\! 0$. Then $\langle g_t,q_t\rangle \!\ge\! \chi_t\|g_t\|_2$, $\chi_t \!=\! \frac{\mu \!-\! \zeta_t \Lambda} {\Lambda(1 \!+\! \zeta_t \!+\! \tau_t)} \!>\! 0$. In particular, under Corollary~\ref{cor:momentum_descent}, one may take $\zeta_t\!=\!2\alpha(1\!+\!\xi_t)\!+\!\xi_t$.
    \end{corollary}
    Corollary~\ref{cor:normalized_alignment} closes the alignment loop for the normalized algorithm. If momentum and polar mixing keep $d_t$ close enough to the damped direction $u_t$, then the normalized direction $q_t$ satisfies Assumption~\ref{assump:align} with an explicit positive constant. Thus, the descent and stationarity results below apply to the actual normalized DAQNM update generated by Algorithm~\ref{alg:daqnm}.

    We next analyze the normalized distance-aware update $\theta_{t+1}$. Since $q_t \!=\! \frac{d_t}{\|d_t\|_2 \!+\! \epsilon}$ we have $\|q_t\|_2\leq1$.

    \begin{theorem}[Descent of normalized distance-aware QNM]
    \label{thm:da_descent}
        Suppose Assumptions~\ref{assump:smooth} and \ref{assump:align} hold. Let $q_t$ be a normalized \DAQNM direction. Then $C(\theta_{t+1}) \!\leq\! C(\theta_t) \!-\! r_t\chi_t\|g_t\|_2 \!+\! \frac{L}{2}r_t^2$.
        In particular, the update is descending whenever $0 \!<\! r_t \!<\! \frac{2\chi_t\|g_t\|_2}{L}$.
    \end{theorem}

    Theorem~\ref{thm:da_descent} explains why normalized optimizers require explicit radius control: once the direction is normalized, the displacement length is governed by $r_t$, not by the original natural-gradient magnitude. 
    It gives a one-step descent condition for an abstract normalized \DAQNM update. We now connect this condition to the specific radius generated by Algorithm~\ref{alg:daqnm}. The next result shows that, under uniform alignment, bounded gradients, and a momentum-gradient comparability condition, the generated radius is both small enough for descent and large enough to yield an $O(1/T)$ stationarity bound.

    \begin{theorem}[Stationarity of normalized \DAQNM with generated radius]
    \label{thm:daqnm_generated_radius_stationarity}
        Suppose Assumptions~\ref{assump:smooth}, \ref{assump:align}, and \ref{assump:lower_bound} hold. Assume there exist constants $\chi\!>\!0$, $G_{\max}\!<\!\infty$, and $0\!<\!a_m \!\le\! b_m\!<\!\infty$ such that, for all $t$, $\langle g_t, q_t\rangle \!\ge\! \chi\|g_t\|_2$, $\|g_t\|_2 \!\le\! G_{\max}$ and $a_m\|g_t\|_2 \!\le\! \|m_t\|_2 \!\le\! b_m\|g_t\|_2$. Let $\eta>0$, $\Delta_0>0$, $\rho\ge0$, and $R_{\max}>0$, and let \DAQNM use the generated radius $r_t$ in (\ref{eq:radius_revised}). If $\eta b_m \!\le\! \frac{\chi}{L}$, $\Delta_0 \!\ge\! \eta a_mG_{\max}$, and $R_{\max}\ge \eta a_mG_{\max}$ then $\eta a_m\|g_t\|_2 \!\le\! r_t \!\le\! \eta b_m\|g_t\|_2 \!\le\! \frac{\chi}{L}\|g_t\|_2$.
        Consequently, the normalized update satisfies $C(\theta_{t+1}) \!\le\! C(\theta_t) \!-\! \frac{\chi \eta a_m}{2} \|g_t\|_2^2$.
        Therefore, for every $T\ge1$, 
        \begin{align} 
        \label{eq:generated_radius_stationarity} 
            \frac{1}{T} \sum_{t=0}^{T-1} \|\nabla C(\theta_t)\|_2^2 \le \frac{ 2\big(C(\theta_0)-C_\star\big) }{ \chi\eta a_m T }. 
        \end{align}
    \end{theorem}
    Theorem~\ref{thm:daqnm_generated_radius_stationarity} is the main stationarity guarantee for the actual normalized \DAQNM update. It connects the algorithmic radius rule directly to descent and stationarity, whereas the fixed-scale stationarity result in Appendix~\ref{app:fixed_scale_stationarity} applies to the auxiliary update $\theta_{t+1} \!=\! \theta_t \!-\! \gamma d_t$.
    Finally, we quantify robustness to gradient and QFIM estimation error.
    \begin{theorem}[Natural-gradient perturbation under QFIM and gradient error]
    \label{thm:perturb_qfim}
       Let $\widehat{\mathbf G} \!=\! \widehat{\mathbf F} \!+\! \lambda \mathbf I$ be a symmetric estimate of $\mathbf G =\mathbf F+\lambda\mathbf I\succeq\mu\mathbf I$. Suppose $\|\widehat g-g\|_2 \!\leq\! \delta_g$, $\|\widehat{\mathbf G}\!-\!\mathbf G\|_2 \!\leq\! \delta_F < \mu$. Let $u \!=\! \mathbf G^{-1}g$, $\widehat u \!=\! \widehat{\mathbf G}^{-1}\widehat g$. Then, $\|\widehat u \!-\! u\|_2 \!\leq\! \frac{\delta_g}{\mu \!-\! \delta_F} \!+\! \frac{\delta_F} {\mu(\mu \!-\! \delta_F)} \|g\|_2$.
    \end{theorem}

    Theorem~\ref{thm:perturb_qfim} motivates damping and structured QFIM approximations. Without damping, small eigenvalues of $\mathbf G$ can amplify gradient and metric-estimation error. With damping, inverse sensitivity is controlled by $\mu$, and hence at least by the damping level $\lambda$.

    Overall, the analysis justifies \DAQNM as a geometry-matched optimizer: QFIM preconditioning calibrates directions to local quantum-state geometry, polar mixing performs spectral direction selection in circuit-semantic blocks, and distance-aware scaling controls the displacement after normalization.
\section{Experiments}
\label{sec:experiments}
\vspace{-0.25cm}
\noindent
We evaluate QNM on VQE, weighted QAOA-MaxCut, and supervised quantum
classification through component ablations that isolate Euclidean
adaptation, matrix spectral normalization, quantum Fisher/QGT
preconditioning (including empirical supervised geometry),
circuit-semantic block choice, and distance-aware radius control. We
compare Adam, Muon, distance-aware Muon (\DAMuon), QNG, and QNM variants
as appropriate. Muon isolates spectral normalization without quantum
geometry, \DAMuon additionally tests radius control, QNG isolates
quantum-geometric preconditioning, and QNM tests spectral normalization
after such preconditioning.

\noindent
For VQE, we report best energy error $E(\theta)-E_\star$. For weighted
QAOA-MaxCut, with achieved expected cut $V_m$ and exact optimum $C_\star$,
we report approximation ratio $V_m/C_\star$, optimality gap
$\Delta_m=C_\star-V_m$, and relative gap reduction
$\rho_{m\leftarrow b}=(\Delta_b-\Delta_m)/\Delta_b$. For supervised
classification, we report mean test accuracy with standard errors.
Finite-shot results are deferred to Appendix~\ref{app:finite_shot_results}
as empirical robustness evidence rather than a probabilistic shot-noise
guarantee.

\noindent
To test the predicted spectral mechanism, we track the active condition
number
$\kappa_{\mathrm{act}}(\mathbf M)=\sigma_1(\mathbf M)/\sigma_r(\mathbf M)$
and effective rank
$r_{\mathrm{eff}}(\mathbf M)=\exp(-\sum_i p_i\log p_i)$, where
$p_i=\sigma_i/\sum_j\sigma_j$, of structured momentum blocks. These
diagnostics measure whether polar mixing reduces active spectral imbalance
by lifting weak singular directions, as predicted by the spectral-floor
analysis.

\subsection{VQE: Layer-Wise Spectral Structure}
\label{subsec:vqe}
\vspace{-0.2cm}
\noindent
We consider VQE for a periodic transverse-field Ising Hamiltonian using a
four-layer hardware-efficient ansatz. Because the ansatz reaches near-exact
energies, the best error $E(\theta)-E_\star$, with $E_\star$ obtained by
classical diagonalization, provides a controlled optimizer diagnostic.
\noindent
As Table~\ref{tab:vqe} shows, QNM improves the best energy error on all
three seeds and reduces the mean error by $80.3\%$. QNM-layer is selected
on all three seeds, supporting the role of circuit-semantic block choice:
layer-wise reshaping organizes parameters across qubits and rotation axes
according to the ansatz structure, allowing polar mixing to regularize the
spectrum of QFIM-preconditioned momentum in an aligned block geometry.
\begin{table}[h]
\centering
\small
\setlength{\tabcolsep}{2pt}
\caption{Four-layer TFIM VQE. ``Best QNM'' selects among layer-, gate-,
and global-block variants; QNM-layer is selected for all seeds.}
\label{tab:vqe}
\begin{tabular}{cccc}
\toprule
Seed & Adam error & Best QNM error & Variant \\
\midrule
0 & $4.832\times10^{-7}$ & $1.529\times10^{-7}$ & QNM-layer \\
1 & $5.142\times10^{-7}$ & $1.509\times10^{-7}$ & QNM-layer \\
2 & $1.218\times10^{-6}$ & $1.334\times10^{-7}$ & QNM-layer \\
\midrule
Mean & $7.385\times10^{-7}$ & $1.457\times10^{-7}$ & -- \\
\bottomrule
\end{tabular}
\end{table}

\subsection{Weighted QAOA: Gate-Structured QNM}
\label{subsec:qaoa}

\noindent
We evaluate generalized QAOA on weighted MaxCut with $n=10$ nodes,
$|E|=23$ edges, depth $p=3$, and exact optimum
$C_\star=17.818521$, computed by exhaustive enumeration. The cost and
mixer parameters form $\boldsymbol{\Gamma}\in\mathbb R^{3\times23}$ and
$\mathbf B\in\mathbb R^{3\times10}$, indexed by depth--edge and
depth--node, respectively. Heterogeneous edge weights directly affect the
cost generators and can induce anisotropy across the edge-indexed
$\boldsymbol{\Gamma}$ update, making this a natural test of semantic block
choice. We use this instance as a controlled component diagnostic rather than a
broad QAOA benchmark.

\noindent
Table~\ref{tab:qaoa_weighted} isolates the three components. QNG reduces
Adam's optimality gap by $43.0\%$, showing the benefit of quantum geometry.
QNM-gate increases this reduction to $96.9\%$, whereas QNM-layer/global
remain near QNG, showing that spectral normalization depends critically on
circuit-semantic block choice. $\DAQNM$-gate further reduces the QNM-gate
gap by $5.7\%$, while \DAMuon performs worse than Adam; neither Euclidean
spectral normalization nor distance-aware scaling alone explains the gain.
The dominant improvement therefore comes from gate-aligned spectral
normalization after QFIM preconditioning, with radius control providing a
smaller additional benefit.
\begin{table}[h]
\centering
\small
\setlength{\tabcolsep}{2pt}
\caption{Weighted generalized QAOA-MaxCut results.}
\label{tab:qaoa_weighted}
\begin{tabular}{lcccc}
\toprule
Method & $V_m$ & $V_m/C_\star$ & $\Delta_m$ \\
\midrule
Adam & $17.355984$ & $0.974042$ & $0.4625$ \\
Muon & $17.094212$ & $0.959351$ & $0.7243$ \\
\DAMuon & $17.098026$ & $0.959565$ & $0.7205$  \\
QNG & $17.555105$ & $0.985217$ & $0.2634$ \\
QNM-layer & $17.535328$ & $0.984107$ & $0.2832$\\
QNM-global & $17.547830$ & $0.984808$ & $0.2707$ \\
QNM-gate & $17.804015$ & $0.999186$ & $0.01451$ \\
$\DAQNM$-gate & $\mathbf{17.804840}$ & $\mathbf{0.999232}$ &
$\mathbf{0.01368}$ \\
\bottomrule
\end{tabular}
\end{table}

\begin{figure}[h]
\centering
\includegraphics[width=0.75\columnwidth]
{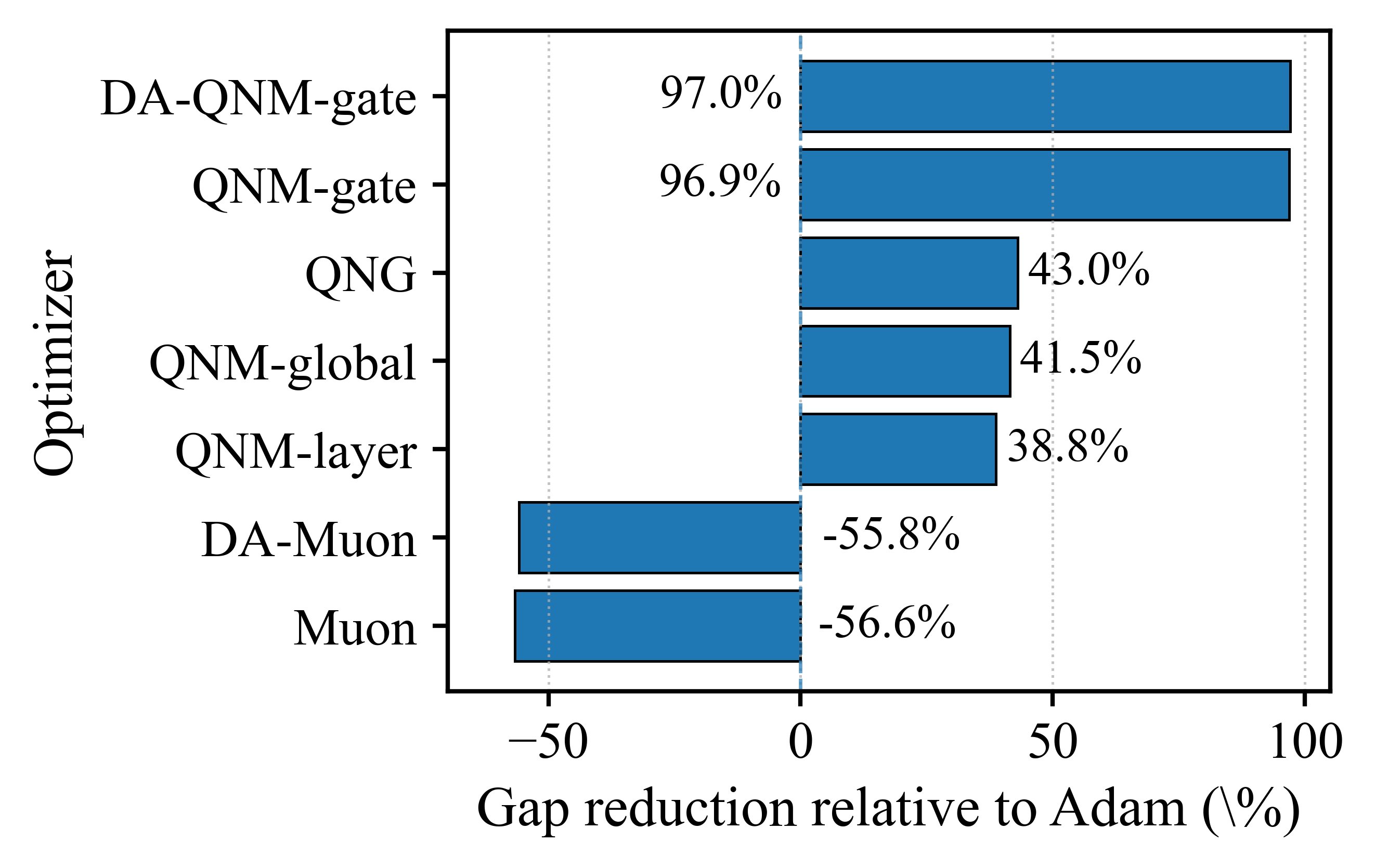}
\caption{Optimality-gap reduction relative to Adam on weighted
QAOA-MaxCut. QNM-gate and DA-QNM-gate achieve $96.9\%$ and $97.0\%$
reductions, respectively.}
\label{fig:qaoa_gap_reduction}
\end{figure}

\noindent
Figure~\ref{fig:qaoa_gap_reduction} summarizes the ablation: Muon/\DAMuon
worsen the Adam gap, QNG improves it, and the largest gains occur only when
QFIM preconditioning is followed by gate-structured spectral normalization.

\subsection{Mechanism Diagnostics: Spectral Regularization}
\label{subsec:diagnostics}

\begin{figure}[h]
\centering
\includegraphics[
    width=0.75\columnwidth,
]
{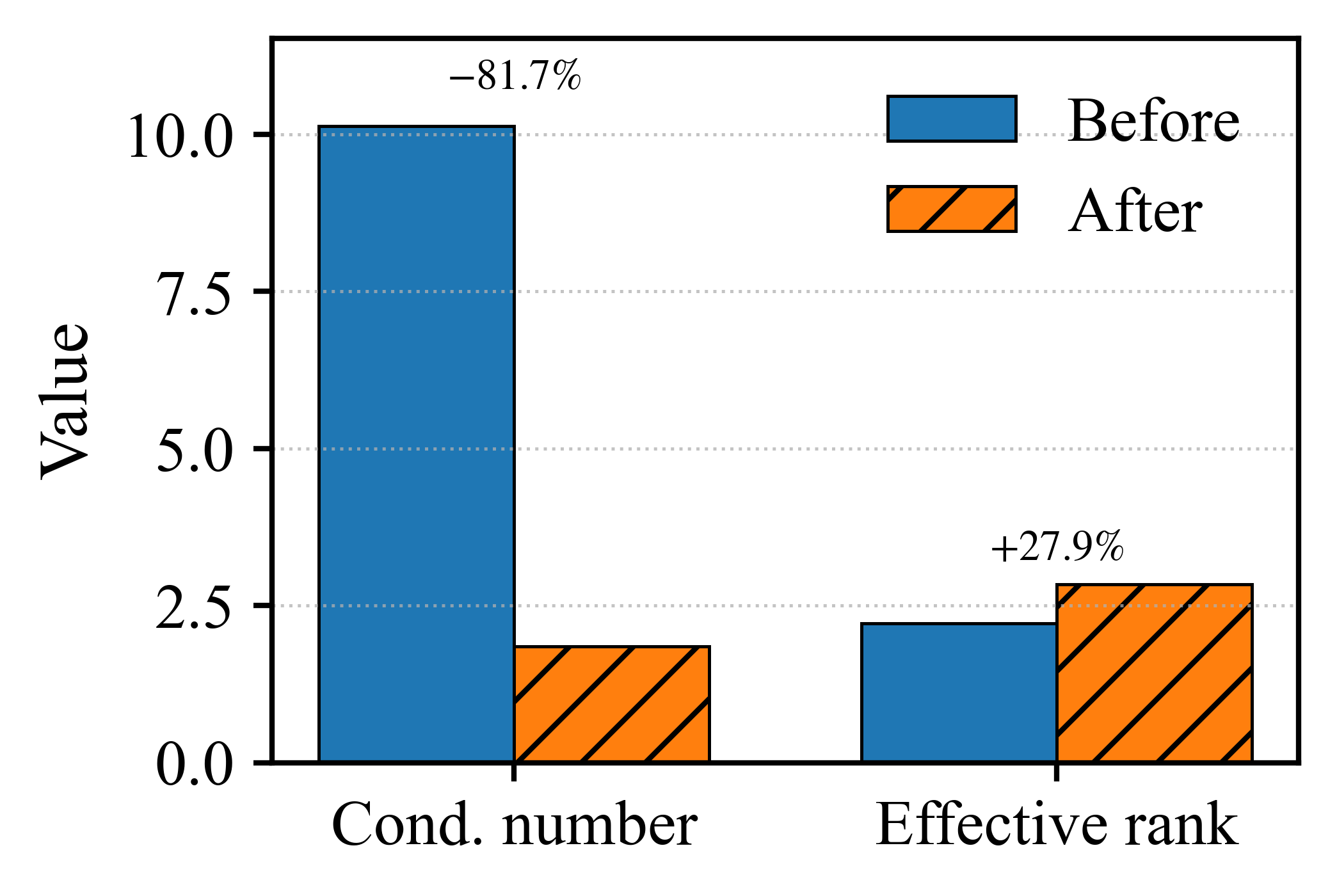}
\caption{Polar mixing reduces the active condition number from $10.13$ to
$1.85$ and increases effective rank from $2.22$ to $2.84$.}
\label{fig:spectral_summary}
\end{figure}

\noindent
We test whether QNM exhibits the mechanism predicted by the spectral
analysis. After QFIM preconditioning, a structured natural-gradient
momentum block can remain spectrally anisotropic; polar mixing should lift
weak active singular directions, reduce spectral imbalance, and increase
effective rank while preserving the active singular subspaces. Figure~\ref{fig:spectral_summary} shows an $81.7\%$ reduction in active
condition number and a $27.9\%$ increase in effective rank, consistent with
the spectral-floor mechanism in Section~\ref{sec:theory}. Thus, the QAOA
gain is accompanied by the predicted reduction in spectral concentration
of the QFIM-preconditioned structured update, complementing the component
and block-structure ablations in Table~\ref{tab:qaoa_weighted}.

\noindent

\subsection{Supervised Quantum Classification}
\label{subsec:classification}

\noindent
We test whether QNM transfers to supervised quantum learning using two
binary Iris tasks: setosa versus rest and the harder versicolor versus
virginica. We use a two-qubit data-reuploading classifier with a
$2\times2\times3$ parameter tensor organized by layer, qubit, and rotation
axis, together with an empirical supervised QGT/QFIM. Muon provides the Euclidean spectral baseline; Diag-QNM uses
diagonal quantum-geometric preconditioning before spectral normalization,
while \DAQNM additionally applies distance-aware radius control. Data
preprocessing, circuit details, and geometry construction are given in
Appendix~\ref{app:classification_details}.


\begin{table}[t]
\centering
\small
\setlength{\tabcolsep}{3pt}
\caption{Iris classification after $120$ steps: mean test accuracy
$\pm$ standard error over three train--test splits.}
\label{tab:iris_120steps}
\begin{tabular}{lcc}
\toprule
Method & Setosa vs. rest & Versicolor vs. virginica \\
\midrule
Adam
& $0.9111\pm0.0128$
& $0.8222\pm0.0222$ \\
Muon
& $0.9407\pm0.0148$
& $0.8333\pm0.0385$ \\
Diag-QNM
& $\mathbf{0.9481\pm0.0196}$
& $0.8556\pm0.0401$ \\
\DAQNM
& $0.9407\pm0.0148$
& $\mathbf{0.8762\pm0.0222}$ \\
\bottomrule
\end{tabular}
\end{table}

\noindent
Table~\ref{tab:iris_120steps} shows that quantum-geometric QNM variants
improve mean accuracy over Adam on both tasks. Diag-QNM performs best on
setosa versus rest ($0.9481$ versus $0.9111$), while \DAQNM performs best
on the harder versicolor-versus-virginica task ($0.8762$ versus $0.8222$),
providing a controlled transfer test of the spectral-natural principle
beyond VQE and QAOA. We do not interpret this small Iris study as a broad
supervised-QML benchmark. Finite-shot results are reported in
Appendix~\ref{app:finite_shot_results} as empirical robustness evidence,
not a probabilistic shot-noise guarantee.

\section{Conclusion}

\vspace{-0.2cm}
\noindent
We introduced \emph{Quantum Natural Muon} (QNM), a structured spectral
natural-gradient framework that applies circuit-semantic polar mixing after
quantum Fisher calibration. We showed that QFIM preconditioning and polar
normalization do not commute, and that Frobenius-matched polar mixing raises
weak active singular directions and improves active spectral conditioning
while remaining a controlled perturbation of QNG. For the distance-aware
realization, we established descent and nonconvex stationarity guarantees.
Across VQE, weighted QAOA, and supervised quantum classification, the
strongest gains occur when spectral blocks align with circuit structure,
supporting the joint use of quantum-state and circuit-structured update
geometry. Future work will study scalable QFIM approximations, finite-shot
theory, and larger hardware-noisy settings.

\newpage
\bibliographystyle{plainnat}
\bibliography{References}

\clearpage
\appendix
\thispagestyle{empty}

\onecolumn
\aistatstitle{Supplementary Materials}

\section{Additional Details of Preliminaries}
\label{app:preliminary}
    This appendix provides a detail of the two geometric structures that motivate \DAQNM. The first is the information geometry of the quantum state manifold, captured locally by QFIM. The second is the matrix or tensor structure induced by the circuit parameterization, such as layer-by-qubit blocks in hardware-efficient VQE ansatzes or depth-by-edge blocks in generalized QAOA. Existing optimizers typically exploit one of these structures but not both.

\subsection{Parameterized Quantum Models}
\label{app:pqms}
    Let $U(\theta)$ be a parameterized quantum circuit acting on $n$ qubits, with parameter vector $\theta\in\mathbb{R}^d$. For a fixed initial state $\ket{0}^{\otimes n}$, the circuit prepares
    \begin{align}
        \ket{\psi(\theta)}
        =
        U(\theta)\ket{0}^{\otimes n}.
    \end{align}
    For Hamiltonian objectives, such as those arising in variational quantum eigensolvers, the loss is
    \begin{align}
    \label{app:hamiltonian_objective}
        C(\theta)
        =
        \bra{\psi(\theta)}\mathbf H\ket{\psi(\theta)},
    \end{align}
    where $\mathbf H$ is a Hermitian problem Hamiltonian \citep{peruzzo2014variational, cerezo2021variational}. In QAOA, $\mathbf H$ is typically a cost Hamiltonian encoding a combinatorial objective \citep{farhi2014quantum, hadfield2019from}. In quantum learning, an input $x$ is encoded through an input-dependent circuit $U(x,\theta)$, measured to produce a feature vector $z(x,\theta)$, and passed to a classical readout $f_{\theta,w}(x) = \sigma\!\left(w^\top z(x,\theta)+b\right)$ with empirical risk
    \begin{align}
    \label{app:supervised_loss}
        \mathcal{L}(\theta,w)
        =
        \frac{1}{N}
        \sum_{i=1}^{N}
        \ell\!\left(f_{\theta,w}(x_i),y_i\right).
    \end{align}
    This setting includes variational quantum classifiers, quantum feature-map models, and data-reuploading classifiers \citep{schuld2019quantum, havlicek2019supervised, benedetti2019parameterized, perezsalinas2020data}.

    Although these objectives differ, they share a structural property: the trainable quantum parameters are not arbitrary unstructured coordinates. A hardware-efficient ansatz may have tensor-shaped parameters $\boldsymbol{\Theta} \!\in\! \mathbb{R}^{L\times n\times r}$ where $L$ is the number of layers, $n$ is the number of qubits, and $r$ is the number of trainable rotation axes per qubit \citep{kandala2017hardware, benedetti2019parameterized}. Generalized QAOA for a graph $G \!=\! (V,E)$ may use separate cost and mixer parameter matrices $\boldsymbol{\Gamma} \!\in\! \mathbb{R}^{p\times |E|}$, $\mathbf B \!\in\! \mathbb{R}^{p\times |V|}$ where $p$ is the QAOA depth \citep{farhi2014quantum, hadfield2019from}. Data-reuploading classifiers similarly organize parameters by reuploading layer, qubit, and rotation family \citep{perezsalinas2020data}.

    We formalize this structure through circuit-semantic reshape maps. Let $\mathcal{S} \!=\! \{1,\ldots,S\}$ index a collection of parameter blocks. For each block $s \!\in\! \mathcal{S}$, let $\mathcal{I}_s \!\subseteq\! \{1,\ldots,d\}$ denote the corresponding parameter indices, and let $\theta^{(s)} \!=\! \theta_{\mathcal{I}_s} \!\in\! \mathbb{R}^{d_s}$ be the associated parameter subvector. We define a reshape operator
    \begin{align}
    \label{app:reshape_map}
        \mathcal{R}_s: \mathbb{R}^{d_s} \rightarrow \mathbb{R}^{a_s\times b_s}, \quad 
        \mathbf M^{(s)} = \mathcal{R}_s(\theta^{(s)}).
    \end{align}
    Examples include layer-wise, gate-family, QAOA cost-angle, QAOA mixer-angle, and data-reuploading reshape maps. These maps encode the circuit-level modeling semantics. A central idea of this paper is that optimizer design should exploit this structure rather than treating $\theta$ as a featureless vector.

\subsection{Quantum Natural Gradient}
\label{app:qng}
    The Euclidean norm of a parameter perturbation does not generally reflect the induced change in the quantum state. Natural-gradient methods address this mismatch by replacing Euclidean geometry with an information-geometric metric \citep{amari1998natural}. In variational quantum models, the corresponding metric is given by the quantum geometric tensor, whose real part defines the QFIM for pure states \citep{provost1980riemannian, stokes2020quantum}. For pure states, define QFIM as
    \begin{align}
    \label{app:qfim}
        [\mathbf F(\theta)]_{ij} 
        &= \operatorname{Re} \big[ \braket{\partial_i\psi(\theta)}{\partial_j\psi(\theta)} \\
        &\quad - \braket{\partial_i \psi(\theta)}{\psi(\theta)} \braket{\psi(\theta)}{\partial_j \psi(\theta)} \big]. \nonumber
    \end{align}
    With this convention, the local infidelity satisfies
    \begin{align}
    \label{app:qfim_local_metric}
        1-\left|\braket{\psi(\theta)}{\psi(\theta+\delta)}\right|^2 = \delta^\top \mathbf F(\theta)\delta + o(\|\delta\|_2^2),
    \end{align}
    up to convention-dependent constant factors used in the literature. Thus, $\mathbf F(\theta)$ measures the local sensitivity of the prepared quantum state to parameter perturbations. Let $g_t \!=\! \nabla C(\theta_t)$ denote the Euclidean gradient of the objective with respect to the quantum parameters, and let $\mathbf F_t \!=\! \mathbf F(\theta_t)$ be the QFIM at the current iterate. Given damping $\lambda>0$, define $\mathbf G_t \!=\! \mathbf F_t \!+\! \lambda \mathbf I$. The damped natural-gradient vector is
    \begin{align}
    \label{app:natural_gradient_vector}
        u_t = \mathbf G_t^{-1}g_t.
    \end{align}
    Equivalently, the corresponding descent displacement is the solution of the local quadratic model
    \begin{align}
    \label{app:qng_variational}
        d_t^{\mathrm{QNG}} \!= \argmin_{d\in\mathbb{R}^d} \Big\{ \langle g_t,d\rangle + \frac{1}{2}d^\top \mathbf G_t d \Big\}  = -\mathbf G_t^{-1}g_t. 
    \end{align}
    Therefore, when the update is written as
    \begin{align}
        \theta_{t+1} = \theta_t-\eta u_t,
    \end{align}
    $u_t$ is the natural-gradient vector, while $-u_t$ is the corresponding descent displacement.

    QNG therefore solves a local state-metric problem: it calibrates the update to the geometry of the quantum state manifold \citep{stokes2020quantum}. However, (\ref{app:natural_gradient_vector}) still produces an unstructured vector. It does not specify how update energy should be distributed across circuit-induced matrix blocks $\mathcal R_s(u_t^{(s)})$. Thus, even after QFIM preconditioning, a reshaped natural-gradient block may remain spectrally anisotropic and dominated by a few structured directions. This motivates the spectral block normalization introduced in Section~\ref{sec:method}.

\subsection{Muon and Normalized Matrix Updates}
\label{app:muon}
    Muon-style optimizers address a different geometry: the spectral geometry of matrix-valued momentum updates \citep{jordan2024muon, bernstein2025modular}. Let $\mathbf M \in \mathbb{R}^{a\times b}$ be a matrix momentum block with compact singular value decomposition $\mathbf M \!=\! \mathbf U\boldsymbol{\Sigma}\mathbf V^\top$, $\operatorname{rank}(\mathbf M) \!=\! r$. The polar factor is
    \begin{align}
    \label{app:polar_factor}
        \operatorname{Polar}(\mathbf M) = \mathbf U\mathbf V^\top.
    \end{align}
    It preserves the active left and right singular subspaces of $\mathbf M$, while setting all nonzero singular values to one:
    \begin{align}
        \sigma_1(\operatorname{Polar}(\mathbf M)) = \cdots = \sigma_r(\operatorname{Polar}(\mathbf M)) = 1.
    \end{align}
    Thus, polar normalization controls matrix-level spectral anisotropy. If $\mathbf M$ is dominated by a few singular values, $\operatorname{Polar}(\mathbf M)$ increases the relative contribution of weaker active directions while preserving the active singular subspaces. In practice, Muon approximates this polar factor using a small number of Newton--Schulz iterations, a standard approach for computing matrix polar factors \citep{higham1986computing, higham2008functions, jordan2024muon}.

    The polar factor also has a local optimization interpretation. It is an optimizer of the spectral-norm constrained linear problem
    \begin{align}
    \label{app:polar_steepest}
        \operatorname{Polar}(\mathbf M)
        \in
        \argmax_{\|\mathbf D\|_2\leq 1}
        \langle \mathbf M,\mathbf D\rangle_F,
    \end{align}
    where $\langle \mathbf A,\mathbf B\rangle_F \!=\! \operatorname{tr}(\mathbf A^\top\mathbf B)$. The optimal value is $\|\mathbf M\|_*$, the nuclear norm of $\mathbf M$, by duality between the spectral and nuclear norms. Therefore, polar normalization is not merely a heuristic rescaling: it selects a steepest matrix direction under a spectral-norm trust region.

    However, normalized matrix optimizers introduce a separate scale-selection problem. Once the update direction is normalized, part of the original magnitude information is removed, and the scalar step radius becomes a central algorithmic choice. Recent work on Distance-Aware Muon studies this direction--scale separation and proposes adaptive radius control for normalized optimization \citep{demidovich2026distanceaware}. This issue is especially relevant in variational quantum circuits, where large normalized steps can overshoot near sharp optima and where gradient or QFIM estimates may be noisy. These observations motivate the distance-aware radius used in \DAQNM.

\subsection{The Missing Natural Direction with Structured Spectral Control Geometry}
\label{app:missing_geometry}
    The preceding discussion reveals a gap. QNG corrects the quantum-state geometry but does not control the active singular spectrum of circuit-structured update blocks. Muon controls the active singular spectrum of matrix updates but ignores the quantum-state geometry. Distance-aware Muon improves scalar radius selection for normalized updates but still operates in a classical matrix-update geometry.

    This paper studies the optimizer design problem that arises when both geometries are present. Given a gradient $g_t$, a damped quantum metric $\mathbf G_t= \mathbf F_t+ \lambda \mathbf I$, and circuit-semantic reshape maps $\{\mathcal R_s\}_{s\in\mathcal S}$, we seek an update satisfying three requirements:
    \begin{itemize}[itemsep=2pt,parsep=0pt,topsep=1pt]
        \item \textbf{Quantum-state calibration:} the update should first correct the Euclidean gradient using the local QFIM geometry.
        \item \textbf{Structured spectral balancing:} after quantum preconditioning, the update should reduce spectral anisotropy in semantically meaningful matrix blocks.
        \item \textbf{Distance-aware displacement:} after normalization, the scalar step radius should be controlled explicitly rather than inherited implicitly from a normalized direction.
    \end{itemize}

    This motivates the \DAQNM pipeline developed in Section~\ref{sec:method}. 
    The key design choice is the order
    rather than applying spectral normalization directly to the raw Euclidean gradient. 
    In general, QFIM preconditioning and blockwise polar normalization do not commute. Thus, \DAQNM is not simply an order-free composition of existing optimizers; it applies spectral normalization in the natural-gradient geometry induced by the quantum model and then controls the resulting normalized displacement through a distance-aware radius.
\section{Additional Theory and Proofs}
\label{app:proofs}
    This appendix provides complete proofs for the theoretical statements in Section~\ref{sec:theory}. We use the notation of the main paper. In particular, $g_t=\nabla C(\theta_t)$, $\mathbf F_t$ is the QFIM, $\mathbf G_t=\mathbf F_t+\lambda \mathbf I$, $u_t=\mathbf G_t^{-1}g_t$, and $d_t$ denotes the fixed-scale spectral-natural direction produced before normalization. The normalized direction is
    \begin{align}
        q_t = \frac{d_t}{\|d_t\|_2 + \epsilon}.
    \end{align}

\subsection{More details of Assumptions}
\label{app:assumption}
    \begin{remark}
        Since $\mathbf F_t\succeq 0$, damping guarantees $\mathbf G_t\succeq \lambda\mathbf I$. Thus, one may always take $\mu=\lambda$ as a conservative lower bound, while a larger $\mu$ may be used when a sharper uniform bound is available. The upper bound is natural for finite-dimensional parameterized circuits with bounded generators on compact parameter domains. The condition number $\Lambda/\mu$ controls the sensitivity of the natural-gradient direction to perturbations and determines how much spectral mixing can be tolerated while preserving descent.
    \end{remark}

    Assumption~\ref{assump:block_norm} holds when the blocks form a disjoint partition of the coordinates and each $\mathcal R_s$ is a reshape operation. If overlapping blocks are used, the same analysis can be stated with upper and lower norm-equivalence constants, but we use the norm-preserving case for clarity.

    Assumption~\ref{assump:align} is a standard alignment condition for normalized nonconvex descent. It states that the normalized direction $q_t$ has a positive component along the Euclidean gradient $g_t$, so that the update $\theta_{t+1}=\theta_t-r_tq_t$ is locally descending for sufficiently small radius $r_t$. Later results provide sufficient conditions under which alignment follows from controlled perturbations of the damped natural-gradient direction.

    Assumption~\ref{assump:lower_bound} is only needed for the stationarity result. It is satisfied by Hamiltonian expectation objectives with bounded spectra, finite QAOA cost objectives, and standard nonnegative empirical losses.

\subsection{Proof of Theorem~\ref{thm:spectral_lmo}}
\label{app:proof_spectral_lmo}
     \begin{proof}
        We prove the result in two steps. First, no feasible matrix $\mathbf D$ with $\|\mathbf D\|_2\leq 1$ can achieve objective value larger than $\|\mathbf M\|_*$. By duality between the spectral norm and the nuclear norm, 
        \begin{align} 
            \max_{\|\mathbf D\|_2\leq 1} \langle \mathbf M,\mathbf D\rangle_F = \|\mathbf M\|_*.
        \end{align} 
        It remains to show that the polar factor attains this upper bound. Let $\mathbf M \!=\! \mathbf U\boldsymbol{\Sigma}\mathbf V^\top$ be the compact singular value decomposition of $\mathbf M$, where $\mathbf U\in\mathbb{R}^{a\times r}$, $\mathbf V\in\mathbb{R}^{b\times r}$, $\mathbf U^\top\mathbf U=\mathbf I_r$, $\mathbf V^\top\mathbf V=\mathbf I_r$, and 
        \begin{align} 
            \boldsymbol{\Sigma} = \operatorname{diag}(\sigma_1,\ldots,\sigma_r), \quad \sigma_i>0. 
        \end{align} 
        For any feasible $\mathbf D\!\in\!\mathbb{R}^{a\times b}$ satisfying $\|\mathbf D\|_2 \!\leq\! 1$, we have 
        \begin{align} 
            \langle \mathbf M,\mathbf D\rangle_F 
            &= \operatorname{Tr}(\mathbf M^\top\mathbf D)  = \operatorname{Tr} \left( (\mathbf U\boldsymbol{\Sigma}\mathbf V^\top)^\top \mathbf D \right) \\ 
            &= \operatorname{Tr} \left( \mathbf V\boldsymbol{\Sigma}\mathbf U^\top\mathbf D \right) 
        \end{align} 
        Using cyclic invariance of the trace, 
        \begin{align} 
            \operatorname{Tr} \left( \mathbf V\boldsymbol{\Sigma}\mathbf U^\top\mathbf D \right) = \operatorname{Tr} \left( \boldsymbol{\Sigma}\mathbf U^\top\mathbf D\mathbf V \right).
        \end{align} 
        Define $\mathbf A \!=\! \mathbf U^\top\mathbf D\mathbf V \in \mathbb{R}^{r\times r}$. Then 
        \begin{align} 
            \langle \mathbf M,\mathbf D\rangle_F = \operatorname{Tr}(\boldsymbol{\Sigma}\mathbf A) = \sum_{i=1}^{r}\sigma_i A_{ii}. 
        \end{align} 
        Since $\mathbf U$ and $\mathbf V$ have orthonormal columns, 
        \begin{align} 
            \|\mathbf A\|_2 = \|\mathbf U^\top\mathbf D\mathbf V\|_2 \leq \|\mathbf U^\top\|_2 \|\mathbf D\|_2 \|\mathbf V\|_2 \leq 1 . 
        \end{align} 
        Hence $|A_{ii}|\leq \|\mathbf A\|_2\leq 1$, and therefore $A_{ii}\leq 1$ for every $i$. Since $\sigma_i>0$, 
        \begin{align} 
            \langle \mathbf M,\mathbf D\rangle_F = \sum_{i=1}^{r} \sigma_i A_{ii} \leq \sum_{i=1}^{r} \sigma_i = \|\mathbf M\|_* . 
        \end{align} 
        Now set $\mathbf D_\star = \mathbf U\mathbf V^\top$. Because $\mathbf U$ and $\mathbf V$ have orthonormal columns, the nonzero singular values of $\mathbf U\mathbf V^\top$ are all equal to one. Hence $\|\mathbf D_\star\|_2=1$, so $\mathbf D_\star$ is feasible. Evaluating the objective at $\mathbf D_\star$, 
        \begin{align} 
            \langle \mathbf M,\mathbf D_\star\rangle_F 
            &= \operatorname{Tr} \left( \mathbf M^\top\mathbf U\mathbf V^\top \right) \\ 
            &= \operatorname{Tr} \left( (\mathbf U\boldsymbol{\Sigma}\mathbf V^\top)^\top \mathbf U\mathbf V^\top \right) \\
            &= \operatorname{Tr}(\boldsymbol{\Sigma}) = \sum_{i=1}^{r}\sigma_i = \|\mathbf M\|_* . 
        \end{align} 
        Thus, $\mathbf D_\star = \mathbf U\mathbf V^\top = \operatorname{Polar}(\mathbf M)$ attains the upper bound. Therefore, 
        \begin{align} 
            \operatorname{Polar}(\mathbf M) \in \argmax_{\|\mathbf D\|_2\leq 1} \langle \mathbf M,\mathbf D\rangle_F . 
        \end{align} 
        The optimal value is $\|\mathbf M\|_*$. Finally, if $\mathbf M=0$, then $\langle \mathbf M,\mathbf D\rangle_F = 0$
        for every feasible $\mathbf D$. Hence every $\mathbf D$ satisfying $\|\mathbf D\|_2\leq 1$ is optimal, and the optimal value is $0=\|\mathbf M\|_*$. This completes the proof. 
    \end{proof}

\subsection{Proof of Proposition~\ref{prop:noncommute}}
\label{app:proof_noncommute}
    Consider a single block with reshape map
    $\mathcal R:\mathbb{R}^{d_s} \!\to\! \mathbb{R}^{a\times b}$.
    For a positive definite metric $\mathbf G \!\succ\! 0$ on this block, the induced blockwise preconditioning operator is
    \begin{align}
    \label{eq:block_metric_operator_appendix}
        \mathcal A_{\mathbf G}(\mathbf M) = \mathcal R \left( \mathbf G^{-1} \mathcal R^{-1}(\mathbf M) \right),
    \end{align}
    where $\mathcal R^{-1}$ is vectorization back to the corresponding parameter coordinates. \DAQNM first applies metric preconditioning and then performs polar normalization, $\mathbf M \!\mapsto\! \operatorname{Polar}\!\left(\mathcal A_{\mathbf G}(\mathbf M)\right)$. A Muon-then-QNG alternative would instead polar-normalize first and then apply the metric, $\mathbf M \!\mapsto\! \mathcal A_{\mathbf G}\!\left(\operatorname{Polar}(\mathbf M)\right)$.

    \begin{proof}
        The proposition states that QFIM preconditioning and blockwise polar normalization do not commute in general. 
        It suffices to give a counterexample. This simplified operator is an instance of the blockwise preconditioning map $\mathcal A_{t,s}$ from the main text, restricted to one block with a fixed positive definite metric $\mathbf G$.
        Let $a=b=2$, $d_s=4$, and let $\mathcal R$ be the standard matrix reshape satisfying
        \begin{align}
            \mathcal R^{-1}(\mathbf M) = \operatorname{vec}(\mathbf M).
        \end{align}
        Choose the positive definite matrix $\mathbf A =
            \begin{pmatrix}
                2 & 0\\
                0 & 1
            \end{pmatrix}$.
        Define the block metric preconditioner by $\mathbf G^{-1} = \mathbf I_2\otimes \mathbf A$
        where $\otimes$ denotes the Kronecker product. Since $\mathbf A\succ 0$, we have $\mathbf G^{-1}\succ 0$, and therefore $\mathbf G\succ 0$. Under column-wise vectorization, $\operatorname{vec}(\mathbf A\mathbf M) = (\mathbf I_2\otimes \mathbf A) \operatorname{vec}(\mathbf M)$.
        Hence, the induced blockwise operator satisfies $\mathcal A_{\mathbf G}(\mathbf M) = \mathbf A\mathbf M$.
        Now take $\mathbf M=\mathbf I_2$. Since $\mathbf M=\mathbf I_2$, its polar factor is $\operatorname{Polar}(\mathbf M) = \mathbf I_2$.
        Therefore, the Muon-then-QNG ordering gives
        \begin{align}
        \label{eq:muon_then_qng_counterexample}
            \mathcal A_{\mathbf G} \left( \operatorname{Polar}(\mathbf M) \right) = \mathcal A_{\mathbf G}(\mathbf I_2) = \mathbf A\mathbf I_2 = \mathbf A.
        \end{align}
        On the other hand, the QNG-then-polar ordering yields $\mathcal A_{\mathbf G}(\mathbf M) = \mathbf A$. Because $\mathbf A$ is symmetric positive definite, its polar factor is
        \begin{align}
            \operatorname{Polar}(\mathbf A) = \mathbf A(\mathbf A^\top\mathbf A)^{-1/2} = \mathbf A(\mathbf A^2)^{-1/2} = \mathbf I_2.
        \end{align}
        Consequently,
        \begin{align}
        \label{eq:qng_then_polar_counterexample}
            \operatorname{Polar} \left( \mathcal A_{\mathbf G}(\mathbf M) \right) = \mathbf I_2.
        \end{align}
        Combining (\ref{eq:muon_then_qng_counterexample}) and (\ref{eq:qng_then_polar_counterexample}), we obtain
        \begin{align}
            \operatorname{Polar} \left( \mathcal A_{\mathbf G}(\mathbf M) \right) = \mathbf I_2 \neq \mathbf A = \mathcal A_{\mathbf G} \left( \operatorname{Polar}(\mathbf M) \right).
        \end{align}
        Therefore, metric preconditioning and polar normalization do not commute in general.
    \end{proof}

\subsection{Proof of Proposition~\ref{prop:active}}
\label{app:proof_active}
    \begin{proof}
        Let $\mathbf M \!=\! \mathbf U\boldsymbol{\Sigma}\mathbf V^\top$ be the compact singular value decomposition of $\mathbf M$. The compact polar factor is $\mathbf P \!=\! \mathbf U\mathbf V^\top$. Then $\mathbf P\mathbf P^\top = \mathbf U\mathbf V^\top \mathbf V\mathbf U^\top = \mathbf U\mathbf U^\top$
        and $\mathbf P^\top\mathbf P = \mathbf V\mathbf U^\top \mathbf U\mathbf V^\top = \mathbf V\mathbf V^\top$.
        Both $\mathbf U\mathbf U^\top$ and $\mathbf V\mathbf V^\top$ are orthogonal projectors of rank $r$. Therefore, their nonzero eigenvalues are all equal to one. Since the nonzero singular values of $\mathbf P$ are the square roots of the nonzero eigenvalues of $\mathbf P^\top\mathbf P$, all active singular values of $\mathbf P$ are equal to one. Hence, $\kappa_{\mathrm{act}}(\mathbf P) = \frac{\sigma_1(\mathbf P)}{\sigma_r(\mathbf P)} = 1$.
        This completes the proof.
    \end{proof}

\subsection{Proof of Proposition~\ref{prop:fro}}
\label{app:proof_fro}
    \begin{proof}
        Since $\mathbf P$ has exactly $r$ active singular values and each is equal to one by Proposition~\ref{prop:active},
        \begin{align}
            \|\mathbf P\|_F^2 = \sum_{i=1}^{r}\sigma_i(\mathbf P)^2 = \sum_{i=1}^{r}1 = r.
        \end{align}
        Therefore, $\|\mathbf P\|_F=\sqrt r$. By the definition of $c$,
        \begin{align}
            \|c\mathbf P\|_F = c\|\mathbf P\|_F = \frac{\|\mathbf M\|_F}{\|\mathbf P\|_F} \|\mathbf P\|_F = \|\mathbf M\|_F.
        \end{align}
        Thus, scaling the polar factor by $c$ preserves the Frobenius norm of the original block.
        
        Finally, the singular values of $c\mathbf P$ are $c$ times the singular values of $\mathbf P$. Since every active singular value of $\mathbf P$ equals one, every active singular value of $c\mathbf P$ equals
        \begin{align}
            c = \frac{\|\mathbf M\|_F}{\|\mathbf P\|_F} = \frac{\|\mathbf M\|_F}{\sqrt r}.
        \end{align}
        This proves the proposition.
    \end{proof}

\subsection{Proof of Proposition~\ref{prop:floor}}
\label{app:proof_floor}
    \begin{proof}
        Using the compact SVD of $\mathbf M$ and the definition of $\mathbf P$, we obtain
        \begin{align}
            \mathbf D 
            &= (1-\alpha) \mathbf U\boldsymbol{\Sigma}\mathbf V^\top + \alpha c \mathbf U\mathbf V^\top \\ 
            &= \mathbf U \left( (1-\alpha)\boldsymbol{\Sigma} + \alpha c\mathbf I_r \right) \mathbf V^\top.
        \end{align}
        The middle matrix
        $(1-\alpha)\boldsymbol{\Sigma} + \alpha c\mathbf I_r$
        is diagonal with entries
        $(1-\alpha)\sigma_i+\alpha c,
            i=1,\ldots,r$.
        Because $\alpha\in[0,1]$, $c>0$, and $\sigma_i>0$, these entries are strictly positive. Therefore, this expression is a compact singular value decomposition of $\mathbf D$ with the same active left and right singular vectors $\mathbf U$ and $\mathbf V$. The active singular values of $\mathbf D$ are precisely $\widetilde{\sigma}_i = (1-\alpha)\sigma_i+\alpha c$
        The spectral-floor bound follows immediately: $\widetilde{\sigma}_i = (1-\alpha)\sigma_i+\alpha c \geq \alpha c$.
        This completes the proof.
    \end{proof}

\subsection{Proof of Corollary~\ref{cor:cond}}
\label{app:proof_cond}
    \begin{proof}
        By Proposition~\ref{prop:floor}, the largest and smallest active singular values of $\mathbf D$ are
        \begin{align}
            \widetilde{\sigma}_1 = (1-\alpha)\sigma_1+\alpha c, \quad \widetilde{\sigma}_r = (1-\alpha)\sigma_r+\alpha c.
        \end{align}
        Therefore,
        \begin{align}
           \kappa_{\mathrm{act}}(\mathbf D) = \frac{\widetilde{\sigma}_1}{\widetilde{\sigma}_r} = \frac{(1-\alpha)\sigma_1+\alpha c} {(1-\alpha)\sigma_r+\alpha c}.
        \end{align}
        To prove $\kappa_{\mathrm{act}}(\mathbf D) \!\leq\! \kappa_{\mathrm{act}}(\mathbf M),$ it suffices to show
        \begin{align}
            \frac{(1-\alpha)\sigma_1+\alpha c} {(1-\alpha)\sigma_r+\alpha c} \leq \frac{\sigma_1}{\sigma_r}.
        \end{align}
        Since all denominators are positive, this is equivalent
        \begin{align}
            \big((1-\alpha)\sigma_1+\alpha c\big)\sigma_r \leq \sigma_1 \big((1-\alpha)\sigma_r+\alpha c\big).
        \end{align}
        Expanding both sides yields
        \begin{align}
            (1-\alpha)\sigma_1\sigma_r+\alpha c\sigma_r \leq (1-\alpha)\sigma_1\sigma_r+\alpha c\sigma_1.
        \end{align}
        Canceling the common term $(1-\alpha)\sigma_1\sigma_r$ reduces the inequality to $\alpha c\sigma_r \!\leq\! \alpha c\sigma_1$. This holds because $\alpha\geq0$, $c>0$, and $\sigma_r\leq\sigma_1$. If $\alpha>0$ and $\sigma_1>\sigma_r$, then $\alpha c\sigma_r < \alpha c\sigma_1$, so the condition-number inequality is strict. This completes the proof.
    \end{proof}

    This result gives the structural reason polar mixing can help. It raises weak active singular directions without changing the active singular subspaces. Thus, polar mixing directly reduces spectral collapse in a structured update block. With the Frobenius-preserving choice
    \begin{align}
        c = \frac{\|\mathbf M\|_F} {\|\operatorname{Polar}(\mathbf M)\|_F},
    \end{align}
    this improvement is achieved while preserving the Frobenius-scale budget of the block.
    
\subsection{Proof of Lemma~\ref{lemma:pert}}
\label{app:proof_pert}
    \begin{proof}
        Assume first that $\mathbf M\neq 0$. Since $\mathbf D = (1-\alpha)\mathbf M+\alpha c\mathbf P$,
        it follows that
        \begin{align}
            \|\mathbf D-\mathbf M\|_F 
            &= \alpha\|c\mathbf P-\mathbf M\|_F \\ 
            &\leq \alpha \left( \|c\mathbf P\|_F+\|\mathbf M\|_F \right),
        \end{align}
        where the last step uses the triangle inequality. By the Frobenius-preserving choice of $c$,
        \begin{align}
            \|c\mathbf P\|_F = c\|\mathbf P\|_F = \|\mathbf M\|_F.
        \end{align}
        Substituting this identity yields
        \begin{align}
            \|\mathbf D-\mathbf M\|_F \leq 2\alpha\|\mathbf M\|_F.
        \end{align}
        If $\mathbf M=0$, then by convention $\mathbf D=0$, so
        \begin{align}
            \|\mathbf D-\mathbf M\|_F = 0 \leq 2\alpha\|\mathbf M\|_F.
        \end{align}
        This completes the proof.
    \end{proof}

\subsection{Proof of Theorem~\ref{thm:qnm_descent}}
\label{app:proof_qnm_descent}
    \begin{proof}
        Since $u_t=\mathbf G_t^{-1}g_t$, we have $g_t=\mathbf G_tu_t$. Because $g_t\neq0$ and $\mathbf G_t$ is positive definite, $u_t\neq0$. 
        
        By Assumption~\ref{assump:qfim}, $\mathbf G_t\succeq \mu\mathbf I$, and therefore
        \begin{align}
        \label{eq:qng_positive_alignment}
            \langle g_t,u_t\rangle = u_t^\top \mathbf G_tu_t \geq \mu\|u_t\|_2^2.
        \end{align}
        Write $d_t = u_t+e_t$, and $e_t = d_t-u_t$. By assumption, $\|e_t\|_2\leq\varepsilon_t\|u_t\|_2$.
        Then
        \begin{align}
            \langle g_t,d_t\rangle 
            &= \langle g_t,u_t\rangle + \langle g_t,e_t\rangle \\ 
            &\geq \langle g_t,u_t\rangle - \|g_t\|_2\|e_t\|_2.
        \end{align}
        Using (\ref{eq:qng_positive_alignment}), we obtain
        \begin{align}
        \label{eq:descent_lower_bound_step1}
            \langle g_t,d_t\rangle \geq \mu\|u_t\|_2^2 - \|g_t\|_2\varepsilon_t\|u_t\|_2.
        \end{align}
        Since $g_t=\mathbf G_tu_t$ and $\mathbf G_t\preceq \Lambda\mathbf I$, we have
        \begin{align}
            \|g_t\|_2 = \|\mathbf G_tu_t\|_2 \leq \Lambda\|u_t\|_2.
        \end{align}
        Substituting this bound into (\ref{eq:descent_lower_bound_step1}) gives
        \begin{align}
        \label{eq:descent_alignment_lower_bound}
            \langle g_t,d_t\rangle \geq (\mu-\varepsilon_t\Lambda)\|u_t\|_2^2.
        \end{align}
        If $\varepsilon_t<\mu/\Lambda$, then $\mu-\varepsilon_t\Lambda>0$. Since $u_t\neq0$, (\ref{eq:descent_alignment_lower_bound}) implies $\langle g_t,d_t\rangle>0$.
        
        It remains to show descent of the objective. By $L$-smoothness assumption, we have
        \begin{align}
            C(\theta_t-\gamma d_t) 
            &\leq C(\theta_t) + \langle g_t,-\gamma d_t\rangle + \frac{L}{2}\|\gamma d_t\|_2^2 \\ 
            &= C(\theta_t) - \gamma\langle g_t,d_t\rangle + \frac{L\gamma^2}{2}\|d_t\|_2^2.
        \end{align}
        The right-hand side is strictly smaller than $C(\theta_t)$ whenever $-\gamma\langle g_t, d_t\rangle \!+\! \frac{L\gamma^2}{2}\|d_t\|_2^2 < 0$. Since $\gamma>0$, this condition is equivalent to $\gamma \!<\! \frac{2\langle g_t,d_t\rangle} {L\|d_t\|_2^2}$. Thus every step size satisfying the condition yields
        \begin{align}
            C(\theta_t - \gamma d_t) < C(\theta_t).
        \end{align}
        This completes the proof.
    \end{proof}

\subsection{Proof of Corollary~\ref{cor:polar_descent}}
\label{app:proof_polar_descent}
    \begin{proof}
        In the no-momentum single-block case, the block $\mathbf M$ is obtained by reshaping $u_t$, and $\mathbf D$ is the corresponding polar-mixed block. By Lemma~\ref{lemma:pert},
        \begin{align}
            \|\mathbf D-\mathbf M\|_F \leq 2\alpha\|\mathbf M\|_F.
        \end{align}
        By norm consistency of the reshape,
        \begin{align}
            \|\mathbf M\|_F = \|u_t\|_2, \quad \|\mathbf D-\mathbf M\|_F = \|d_t-u_t\|_2.
        \end{align}
        Therefore,
        \begin{align}
            \|d_t-u_t\|_2\leq2\alpha\|u_t\|_2.
        \end{align}
        Thus, the relative perturbation condition in Theorem~\ref{thm:qnm_descent} holds with $\varepsilon_t\leq 2\alpha$. If $2\alpha<\mu/\Lambda$, then $\varepsilon_t<\mu/\Lambda$. The claim follows directly from Theorem~\ref{thm:qnm_descent}.
    \end{proof}

\subsection{Proof of Corollary~\ref{cor:momentum_descent}}
\label{app:proof_momentum_descent}
    \begin{proof}
        By Lemma~\ref{lemma:pert} applied blockwise, for each block $s\in\mathcal S$, we have
        \begin{align}
            \|\mathbf D_t^{(s)}-\mathbf M_t^{(s)}\|_F \leq 2\alpha\|\mathbf M_t^{(s)}\|_F.
        \end{align}
        Using the norm-consistent block decomposition in Assumption~\ref{assump:block_norm}, we obtain
        \begin{align}
            \|d_t-m_t\|_2^2 
            &= \sum_{s\in\mathcal S} \|\mathbf D_t^{(s)}-\mathbf M_t^{(s)}\|_F^2 \\ 
            &\leq 4\alpha^2 \sum_{s\in\mathcal S} \|\mathbf M_t^{(s)}\|_F^2 = 4\alpha^2\|m_t\|_2^2 .
        \end{align}
        Therefore, we obtain
        \begin{align}
        \label{eq:dt_mt_bound}
            \|d_t-m_t\|_2 \leq 2\alpha\|m_t\|_2 .
        \end{align}
        By the momentum deviation assumption,
        \begin{align}
            \|m_t\|_2 \leq \|u_t\|_2+\|m_t-u_t\|_2 \leq (1+\xi_t)\|u_t\|_2 .
        \end{align}
        Combining this bound with (\ref{eq:dt_mt_bound}) gives
        \begin{align}
            \|d_t-m_t\|_2 \leq 2\alpha(1+\xi_t)\|u_t\|_2 .
        \end{align}
        Finally, by the triangle inequality,
        \begin{align}
            \|d_t-u_t\|_2 &\leq \|d_t-m_t\|_2+\|m_t-u_t\|_2 \\ &\leq \left( 2\alpha(1+\xi_t)+\xi_t \right) \|u_t\|_2 .
        \end{align}
        If the momentum descent condition holds, then the perturbation condition in Theorem~\ref{thm:qnm_descent} is satisfied with $\varepsilon_t=2\alpha(1+\xi_t)+\xi_t$. Therefore, Theorem~\ref{thm:qnm_descent} implies $\langle g_t,d_t\rangle>0$.
        This completes the proof.
    \end{proof}

    The conditions in Corollaries~\ref{cor:polar_descent} and \ref{cor:momentum_descent} are sufficient, not necessary. Corollary~\ref{cor:polar_descent} isolates the no-momentum single-block case, while Corollary~\ref{cor:momentum_descent} connects the analysis to the actual \DAQNM update, where polar mixing is applied to natural-gradient momentum $m_t$. When $\xi_t=0$, the momentum result reduces to the no-momentum perturbation condition $2\alpha<\mu/\Lambda$. When $\xi_t>0$, descent can still be guaranteed provided the accumulated momentum remains close enough to the current damped natural-gradient direction. In practice, substantially larger $\alpha$ may work because the polar perturbation is often aligned with the natural-gradient block. The results nevertheless formalize that the fixed-scale QNM direction continuously interpolates from QNG to a spectrally normalized natural-gradient update while preserving descent for sufficiently small mixing and sufficiently small momentum lag.

\subsection{Proof of Corollary~\ref{cor:normalized_alignment}}
\label{app:proof_normalized_alignment}
    \begin{proof}
        Assume
        \begin{align}
            \|d_t-u_t\|_2\le \zeta_t\|u_t\|_2, \quad \zeta_t<\mu/\Lambda.
        \end{align}
        We write
        \begin{align}
            d_t=u_t+e_t,\quad\|e_t\|_2\le\zeta_t\|u_t\|_2.
        \end{align}
        Since $g_t=\mathbf G_tu_t$, Assumption~\ref{assump:qfim} implies
        \begin{align}
            \langle g_t,u_t\rangle = u_t^\top \mathbf G_tu_t \ge \mu\|u_t\|_2^2, \quad \|g_t\|_2 = \|\mathbf G_tu_t\|_2 \le \Lambda\|u_t\|_2.
        \end{align}
        Therefore, by Cauchy--Schwarz we can obtain
        \begin{align}
            \langle g_t,d_t\rangle = \langle g_t,u_t\rangle+\langle g_t,e_t\rangle \ge \mu\|u_t\|_2^2 - \|g_t\|_2\|e_t\|_2 \ge (\mu-\zeta_t\Lambda)\|u_t\|_2^2.
        \end{align}
        Since $\zeta_t<\mu/\Lambda$, the coefficient $\mu-\zeta_t\Lambda$ is strictly positive. Hence $d_t$ has positive inner product with the gradient.

        Next, we control the denominator introduced by normalization. By the triangle inequality,
        \begin{align}
            \|d_t\|_2 + \epsilon \le \|u_t\|_2 + \|e_t\|_2 + \epsilon \le (1+\zeta_t+\tau_t) \|u_t\|_2,
        \end{align}
        where $\epsilon\le\tau_t\|u_t\|_2$. Since the numerator lower bound is positive, dividing by this upper bound yields a valid lower bound:
        \begin{align}
            \langle g_t,q_t\rangle = \left\langle g_t,\frac{d_t}{\|d_t\|_2+\epsilon}\right\rangle = \frac{\langle g_t,d_t\rangle} {\|d_t\|_2+\epsilon} \ge \frac{\mu-\zeta_t\Lambda} {1+\zeta_t+\tau_t} \|u_t\|_2.
        \end{align}
        Using $\|g_t\|_2\le\Lambda\|u_t\|_2$, we obtain
        \begin{align}
            \langle g_t,q_t\rangle \ge \frac{\mu-\zeta_t\Lambda} {\Lambda(1+\zeta_t+\tau_t)} \|g_t\|_2.
        \end{align}
        Thus Assumption~\ref{assump:align} holds with
        \begin{align}
            \chi_t=\frac{\mu-\zeta_t\Lambda}{\Lambda(1+\zeta_t+\tau_t)}>0.
        \end{align}
        The constant is strictly positive because
        \begin{align}
            \zeta_t<\frac{\mu}{\Lambda} \quad\Longrightarrow\quad \mu-\zeta_t\Lambda>0
        \end{align}
        and the denominator $\Lambda(1+\zeta_t+\tau_t)$ is positive.
        
        It remains only to connect this bound to the momentum-based \DAQNM construction. Under Corollary~\ref{cor:momentum_descent} one may take $\zeta_t=2\alpha(1+\xi_t)+\xi_t$. If this quantity satisfies 
        \begin{align}
            2\alpha(1+\xi_t)+\xi_t<\frac{\mu}{\Lambda}
        \end{align}
        then the normalized \DAQNM direction satisfies the stated alignment bound. This proves the corollary.
    \end{proof}

\subsection{Proof of Theorem~\ref{thm:da_descent}}
\label{app:proof_da_descent}
    \begin{proof}
        By $L$-smoothness, for the update $\theta_{t+1}=\theta_t-r_tq_t$, with displacement $-r_tq_t$,
        \begin{align}
            C(\theta_t \!-\! r_tq_t) 
            &\leq C(\theta_t) \!+\! \langle g_t, -r_tq_t\rangle \!+\! \frac{L}{2}\|-r_tq_t\|_2^2 \\ 
            &= C(\theta_t) \!-\! r_t\langle g_t,q_t\rangle \!+\! \frac{L}{2}r_t^2\|q_t\|_2^2.
        \end{align}
        Using Assumption~\ref{assump:align}, $\langle g_t, q_t\rangle \!\geq\! \chi_t \|g_t\|_2$ and using $\|q_t\|_2\leq1$, we obtain
        \begin{align}
            C(\theta_{t+1}) \leq C(\theta_t) - r_t \chi_t \|g_t\|_2 + \frac{L}{2} r_t^2.
        \end{align}
        The right-hand side is strictly smaller than $C(\theta_t)$ whenever 
        \begin{align}
            -r_t \chi_t\|g_t\|_2 + \frac{L}{2} r_t^2 < 0.
        \end{align}
        Since $r_t>0$, this condition is equivalent to
        \begin{align}
            r_t < \frac{2 \chi_t \|g_t\|_2}{L}.
        \end{align}
        This proves the theorem.
    \end{proof}

    \begin{remark}
        Theorem~\ref{thm:da_descent} explains why normalized optimizers require explicit radius control. Once the direction has been normalized, the displacement length is governed by $r_t$, not by the original magnitude of the natural-gradient vector. The distance-aware radius in \DAQNM is therefore a practical mechanism for controlling the scalar displacement along a structured spectral-natural direction.
    \end{remark}

\subsection{Proof of Theorem~\ref{thm:daqnm_generated_radius_stationarity}}
\label{app:proof_generated_radius_stationarity}
    \begin{proof}
        First, we prove that the radius generated by \DAQNM is gradient-scaled. The generated radius is
        \begin{align}
            r_t = \min \left\{ \eta\|m_t\|_2, \Delta_0+\rho\max_{0\le k\le t}\|\theta_k-\theta_0\|_2, R_{\max} \right\}.
        \end{align}
        By the momentum-gradient comparability assumption,
        \begin{align}
            a_m\|g_t\|_2 \le \|m_t\|_2 \le b_m\|g_t\|_2.
        \end{align}
        Therefore, we obtain
        \begin{align}
            \eta\|m_t\|_2 \le \eta b_m\|g_t\|_2.
        \end{align}
        Since $r_t$ is the minimum of three terms,
        \begin{align}
            r_t\le \eta\|m_t\|_2\le \eta b_m\|g_t\|_2.
        \end{align}
        By the assumed condition $\eta b_m\le\chi/L$,
        \begin{align}
            r_t \le \frac{\chi}{L}\|g_t\|_2.
        \end{align}
        Next, we prove the lower bound. The first argument of the minimum satisfies
        \begin{align}
            \eta\|m_t\|_2 \ge \eta a_m\|g_t\|_2.
        \end{align}
        Since $\|g_t\|_2\le G_{\max}$ and
        $\Delta_0\ge\eta a_mG_{\max}$, we have
        \begin{align}
            \Delta_0+\rho\max_{0\le k\le t}\|\theta_k-\theta_0\|_2 \ge \Delta_0 \ge \eta a_mG_{\max} \ge \eta a_m\|g_t\|_2,
        \end{align}
        where we use $\rho\ge0$. Similarly,
        \begin{align}
            R_{\max} \ge \eta a_mG_{\max} \ge \eta a_m\|g_t\|_2.
        \end{align}
        All three terms inside the minimum are therefore at least $\eta a_m\|g_t\|_2$. Thus, we obtain
        \begin{align}
            r_t\ge \eta a_m\|g_t\|_2.
        \end{align}
        Combining the lower and upper bounds gives
        \begin{align}
            \eta a_m\|g_t\|_2 \le r_t \le \eta b_m\|g_t\|_2 \le \frac{\chi}{L}\|g_t\|_2.
        \end{align}
        By Theorem~\ref{thm:da_descent} with uniform alignment constant $\chi$,
        \begin{align}
            C(\theta_{t+1}) \le C(\theta_t)-r_t\chi\|g_t\|_2+\frac{L}{2}r_t^2.
        \end{align}
        Using $r_t\le(\chi/L)\|g_t\|_2$, we have
        \begin{align}
            \frac{L}{2}r_t^2 \le \frac{1}{2}r_t\chi\|g_t\|_2.
        \end{align}
        Therefore,
        \begin{align}
            C(\theta_{t+1})\le C(\theta_t)-\frac{1}{2}r_t\chi\|g_t\|_2.
        \end{align}
        Using $r_t\ge \eta a_m\|g_t\|_2$, we get
        \begin{align}
            C(\theta_{t+1}) \le C(\theta_t) - \frac{\chi\eta a_m}{2}\|g_t\|_2^2.
        \end{align}
        Summing from $t=0$ to $T-1$ yields
        \begin{align}
            \frac{\chi\eta a_m}{2} \sum_{t=0}^{T-1} \|g_t\|_2^2 \le C(\theta_0)-C(\theta_T).
        \end{align}
        Since $C(\theta_T)\ge C_\star$, we have
        \begin{align}
            \sum_{t=0}^{T-1} \|g_t\|_2^2 \le \frac{2(C(\theta_0)-C_\star)} {\chi\eta a_m}.
        \end{align}
        Dividing by $T$ gives
        \begin{align}
            \frac{1}{T} \sum_{t=0}^{T-1} \|\nabla C(\theta_t)\|_2^2 \le \frac{ 2(C(\theta_0)-C_\star) }{ \chi\eta a_mT }.
        \end{align}
        This proves the theorem.
    \end{proof}

    Theorem~\ref{thm:daqnm_generated_radius_stationarity} closes the gap between the descent condition for an abstract normalized step and the radius actually produced by Algorithm~\ref{alg:daqnm}. The upper momentum-gradient comparability condition ensures that the first term $\eta\|m_t\|_2$ makes the generated radius small enough for descent, while the lower comparability condition and the lower bounds on $\Delta_0$ and $R_{\max}$ prevent the distance-aware and cap terms from making the radius too small to yield a stationarity rate. Thus, the generated \DAQNM radius is both safe and large enough to produce an $O(1/T)$ stationarity bound.

\subsection{Proof of Theorem~\ref{thm:perturb_qfim}}
\label{app:proof_perturb_qfim}
    \begin{proof}
        Let $\mathbf G=\mathbf F+\lambda\mathbf I\succeq\mu\mathbf I, \quad \widehat{\mathbf G} = \widehat{\mathbf F}+\lambda\mathbf I$
        and suppose $\|\widehat{\mathbf G}-\mathbf G\|_2\le\delta_F<\mu$.
        By Weyl's inequality, $\widehat{\mathbf G} \succeq (\mu-\delta_F)\mathbf I$
        and hence
        \begin{align}
            \|\widehat{\mathbf G}^{-1}\|_2 \le \frac{1}{\mu-\delta_F}, \quad \|\mathbf G^{-1}\|_2 \le \frac{1}{\mu}.
        \end{align}
        Let $u=\mathbf G^{-1}g, \quad \widehat u=\widehat{\mathbf G}^{-1}\widehat g$.
        Then
        \begin{align}
            \widehat u-u = \widehat{\mathbf G}^{-1}(\widehat g-g) + \left(\widehat{\mathbf G}^{-1}-\mathbf G^{-1}\right)g.
        \end{align}
        The first term satisfies
        \begin{align}
            \|\widehat{\mathbf G}^{-1}(\widehat g-g)\|_2 \le \frac{\delta_g}{\mu-\delta_F}.
        \end{align}
        For the second term, use the inverse-difference identity
        \begin{align}
            \widehat{\mathbf G}^{-1}-\mathbf G^{-1} = \widehat{\mathbf G}^{-1} (\mathbf G-\widehat{\mathbf G}) \mathbf G^{-1}.
        \end{align}
        Therefore,
        \begin{align}
            \|\widehat{\mathbf G}^{-1}-\mathbf G^{-1}\|_2 \le \frac{1}{\mu-\delta_F} \delta_F \frac{1}{\mu} = \frac{\delta_F}{\mu(\mu-\delta_F)}.
        \end{align}
        Combining the two bounds gives
        \begin{align}
            \|\widehat u-u\|_2 \le \frac{\delta_g}{\mu-\delta_F} + \frac{\delta_F}{\mu(\mu-\delta_F)} \|g\|_2.
        \end{align}
        This proves the theorem.
    \end{proof}

\subsection{Auxiliary Fixed-Scale Stationarity Result}
\label{app:fixed_scale_stationarity}
    The main text analyzes the actual normalized distance-aware update used by \DAQNM, $\theta_{t+1}=\theta_t-r_tq_t$. For completeness, we also record a standard fixed-scale stationarity result for the auxiliary update $\theta_{t+1}=\theta_t-\gamma d_t$. This result clarifies the descent mechanism for fixed-scale spectral-natural directions, but it is not the main stationarity guarantee for the implemented distance-aware algorithm.

    \begin{theorem}[Fixed-scale nonconvex stationarity]
    \label{thm:fixed_scale_stationarity}
        Assume $C$ is $L$-smooth and lower bounded by $C_\star$. Consider updates of the form $\theta_{t+1}=\theta_t-\gamma d_t$. Suppose the update directions satisfy
        \begin{align}
        \label{eq:fixed_scale_direction_conditions}
            \langle g_t,d_t\rangle \ge \nu\|g_t\|_2^2,
            \quad
            \|d_t\|_2 \le B\|g_t\|_2
        \end{align}
        for constants $\nu>0$ and $B<\infty$. If
        \begin{align}
            0<\gamma<\frac{2\nu}{LB^2},
        \end{align}
        then
        \begin{align}
        \label{eq:fixed_scale_stationarity_average_bound}
            \frac{1}{T} \sum_{t=0}^{T-1} \|\nabla C(\theta_t)\|_2^2 \le \frac{ C(\theta_0)-C_\star }{ \gamma\left(\nu-\frac{L\gamma B^2}{2}\right)T } .
        \end{align}
        Consequently,
        \begin{align}
            \min_{0\le t<T}\|\nabla C(\theta_t)\|_2^2 = O\!\Big(\frac{1}{T}\Big).
        \end{align}
    \end{theorem}
    \begin{proof}
        By $L$-smoothness,
        \begin{align}
            C(\theta_{t+1}) &= C(\theta_t-\gamma d_t) \\ &\le C(\theta_t) - \gamma\langle g_t,d_t\rangle + \frac{L\gamma^2}{2}\|d_t\|_2^2 .
        \end{align}
        Using the alignment and norm bounds in (\ref{eq:fixed_scale_direction_conditions}),
        \begin{align}
            C(\theta_{t+1}) &\le C(\theta_t) - \gamma\nu\|g_t\|_2^2 + \frac{L\gamma^2B^2}{2}\|g_t\|_2^2 \\ &= C(\theta_t) - \gamma \Big( \nu-\frac{L\gamma B^2}{2} \Big) \|g_t\|_2^2 .
        \end{align}
        The step-size condition ensures $\nu-\frac{L\gamma B^2}{2}>0$.
        Rearranging gives
        \begin{align}
            \gamma \Big( \nu-\frac{L\gamma B^2}{2} \Big) \|g_t\|_2^2 \le C(\theta_t)-C(\theta_{t+1}).
        \end{align}
        Summing over $t=0,\ldots,T-1$, the right-hand side telescopes:
        \begin{align}
            \gamma \Big( \nu-\frac{L\gamma B^2}{2} \Big) \sum_{t=0}^{T-1}\|g_t\|_2^2 &\le C(\theta_0)-C(\theta_T) \\ &\le C(\theta_0)-C_\star .
        \end{align}
        Dividing by $\gamma \Big( \nu-\frac{L\gamma B^2}{2} \Big)T$
        yields (\ref{eq:fixed_scale_stationarity_average_bound}). Finally,
        \begin{align}
            \min_{0\le t<T}\|\nabla C(\theta_t)\|_2^2 \le \frac{1}{T} \sum_{t=0}^{T-1}\|\nabla C(\theta_t)\|_2^2,
        \end{align}
        so the minimum stationarity measure is also $O(1/T)$.
    \end{proof}
    
    Theorem~\ref{thm:fixed_scale_stationarity} is auxiliary. The main stationarity guarantee for the implemented algorithm is Theorem~\ref{thm:daqnm_generated_radius_stationarity}, which applies to the normalized distance-aware update generated by Algorithm~\ref{alg:daqnm}.

\section{Additional Experimental Details}
\label{app:experiment_details}
   This section provides supplementary implementation details for the experiments in Section~\ref{sec:experiments}. The main text reports the primary numerical tables and mechanism diagnostics, while this appendix records setup details, architecture diagrams, learning-curve evidence, and finite-shot robustness results. These details improve reproducibility and clarify how the optimizer variants are instantiated across VQE, weighted QAOA, and supervised quantum classification.

\subsection{VQE Setup} 
\label{app:vqe_details} 
\begin{figure}[h] 
    \centering 
        \includegraphics[width=0.625\columnwidth]{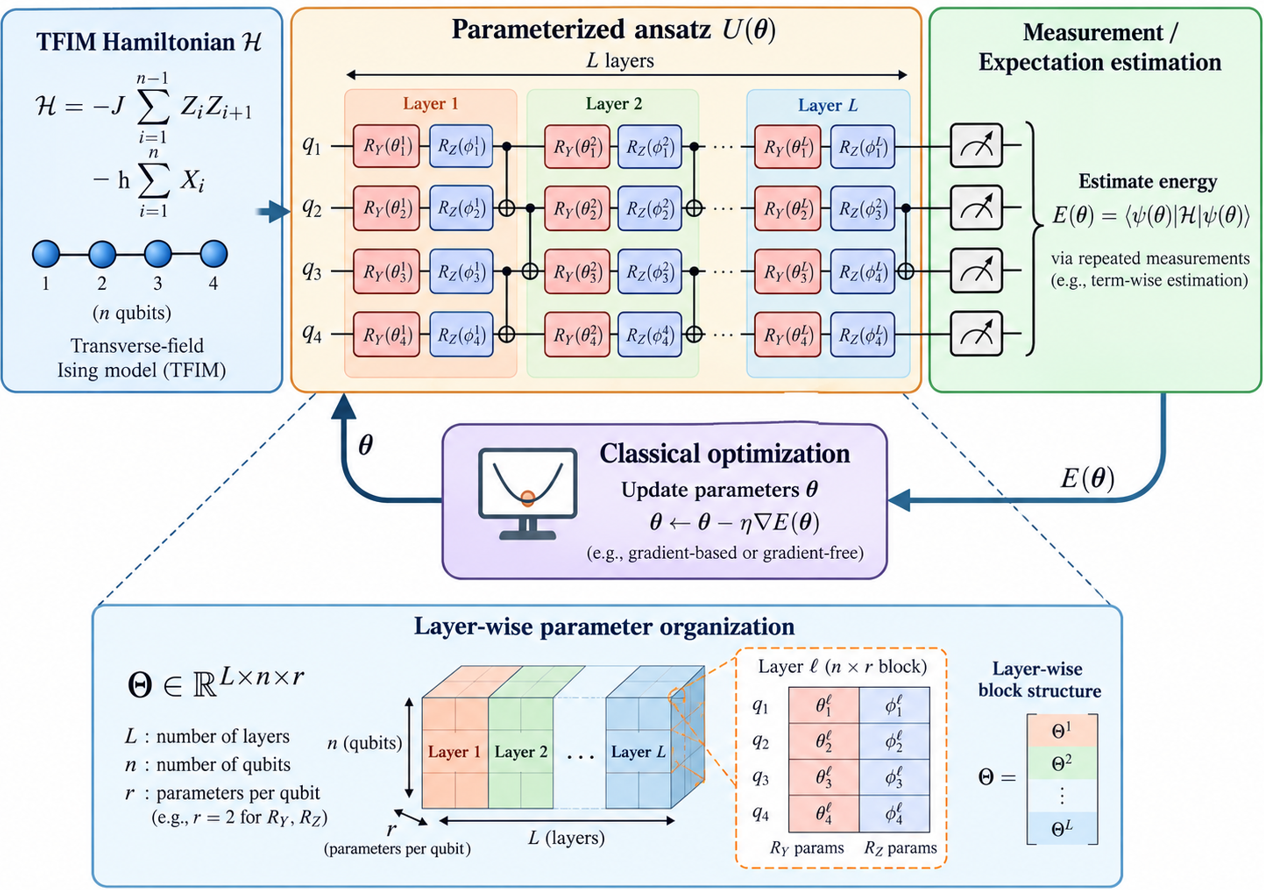} 
        \caption{VQE experimental architecture. A parameterized quantum circuit prepares the trial state $\ket{\psi(\theta)}$, the Hamiltonian expectation $E(\theta)=\bra{\psi(\theta)}\mathbf H\ket{\psi(\theta)}$ is evaluated, and the optimizer updates the circuit parameters. In the \DAQNM variants, gradients are first preconditioned by the damped QFIM, then reshaped into circuit-semantic blocks for spectral normalization and distance-aware radius control.} 
    \label{fig:vqe_model} 
    \end{figure}

    Fig.~\ref{fig:vqe_model} summarizes the VQE experimental workflow used in our simulations, including state preparation, Hamiltonian expectation evaluation, optimizer update, and the \DAQNM-specific QFIM and spectral normalization steps.

    The VQE experiment uses a transverse-field Ising Hamiltonian with periodic boundary conditions. The variational objective is $E(\theta) = \bra{\psi(\theta)}\mathbf H\ket{\psi(\theta)}$.
    The exact ground-state energy $E_\star$ is computed by classical diagonalization, and the reported metric is the best energy error achieved over the optimization trajectory, $\min_t \big(E(\theta_t)-E_\star\big)$.
    We use a four-layer hardware-efficient ansatz. The QNM variants differ only in the circuit-semantic reshape used for spectral normalization: layer-wise blocks group parameters within each variational layer, gate-wise blocks group parameters by gate family, and the global variant applies a single larger reshape to the full parameter vector. The main text reports the best QNM variant per seed; in the current runs, QNM-layer is selected for all three tested seeds. This supports the interpretation that layer-wise spectral structure is the most compatible block geometry for this VQE ansatz.

\subsection{Weighted QAOA Setup}
\label{app:qaoa_details}
\begin{figure}[h]
        \centering
        \includegraphics[width=0.625\columnwidth]{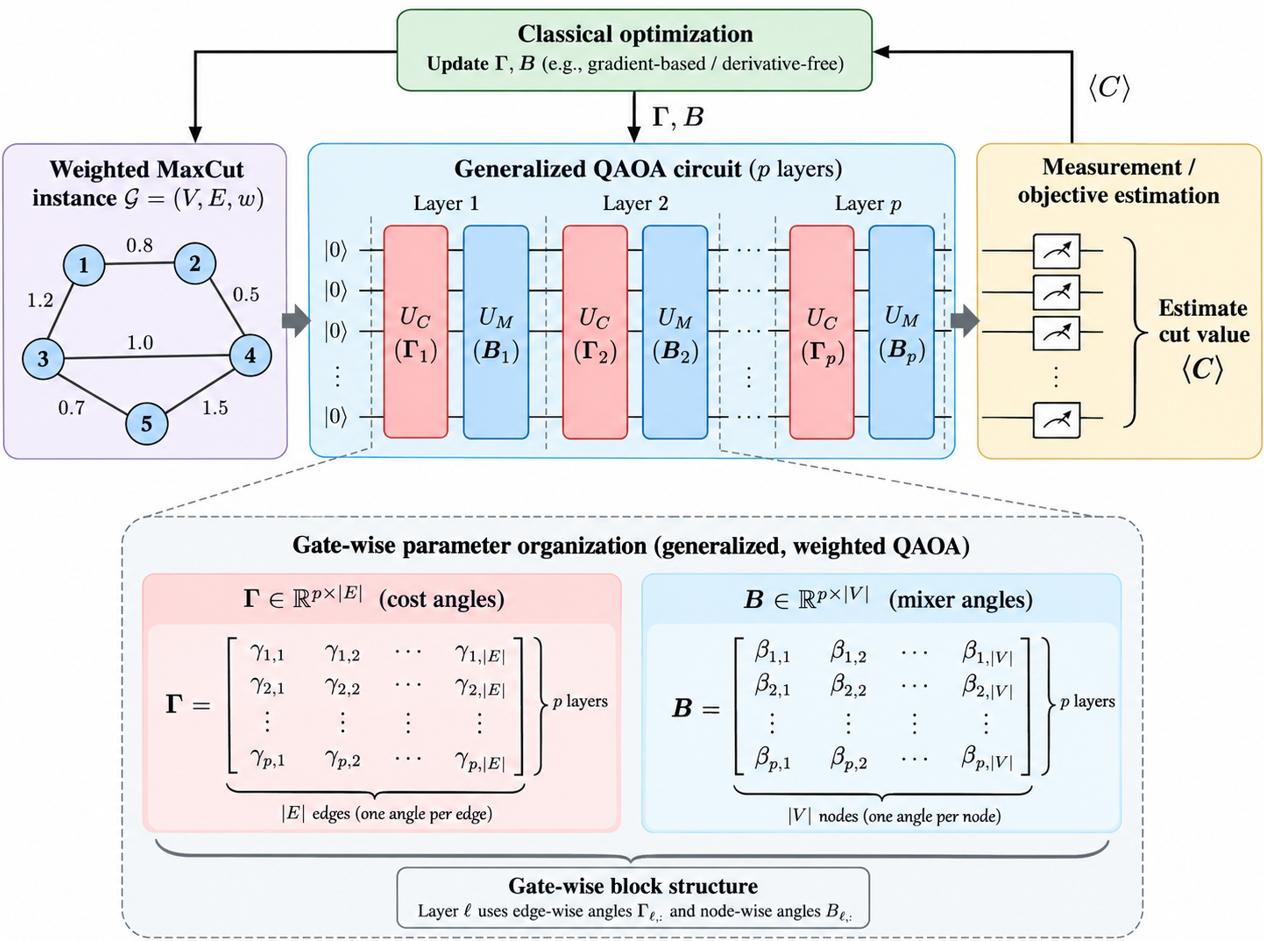}
        \caption{Weighted QAOA experimental architecture. The circuit alternates
        between cost unitaries parameterized by
        $\boldsymbol{\Gamma}\in\mathbb R^{3\times23}$ and mixer unitaries
        parameterized by $\mathbf B\in\mathbb R^{3\times10}$. Gate-structured
        QNM applies damped QFIM preconditioning followed by spectral normalization
        on circuit-semantic blocks aligned with the cost-angle and mixer-angle
        parameter tensors.}
        \label{fig:qaoa_model}
    \end{figure}

    Fig.~\ref{fig:qaoa_model} summarizes the generalized QAOA workflow used in the weighted MaxCut experiment and highlights the parameter blocks on which gate-structured QNM operates.
    
    The weighted QAOA experiment uses a MaxCut instance with $n=10, \quad |E|=23, \quad p=3$
    The exact optimal cut value is computed by exhaustive enumeration: $C_\star = 17.818521$.
    The generalized QAOA parameters decompose into a cost-angle block and a mixer-angle block, $\boldsymbol{\Gamma}\in\mathbb R^{3\times23}, \quad \mathbf B\in\mathbb R^{3\times10}$.
    Here $\boldsymbol{\Gamma}$ is indexed by depth and edge, while $\mathbf B$ is indexed by depth and node. This circuit-semantic decomposition makes weighted QAOA a natural testbed for gate-structured QNM: heterogeneous edge weights affect the cost-unitary generators and can induce anisotropic columns in the natural-gradient update associated with $\boldsymbol{\Gamma}$. All methods are evaluated on the same weighted instance and the same initialization seed. The purpose of this experiment is not to claim broad QAOA dominance from one instance, but to isolate whether QFIM preconditioning, gate-structured spectral normalization, and distance-aware radius control improve optimization on a controlled circuit-structured problem.

\subsection{Additional Mechanism Figures}
\label{app:additional_mechanism_figures}
    This section provides an additional spectral diagnostic supporting the mechanism discussed in Section~\ref{sec:experiments}. The main text reports summary statistics such as the active condition number and effective rank. Here, we show the corresponding active singular spectrum directly, which makes visible how polar mixing changes the geometry of a representative QNM-gate momentum block.

    \begin{figure}[h]
        \centering
        \includegraphics[width=0.45\columnwidth]{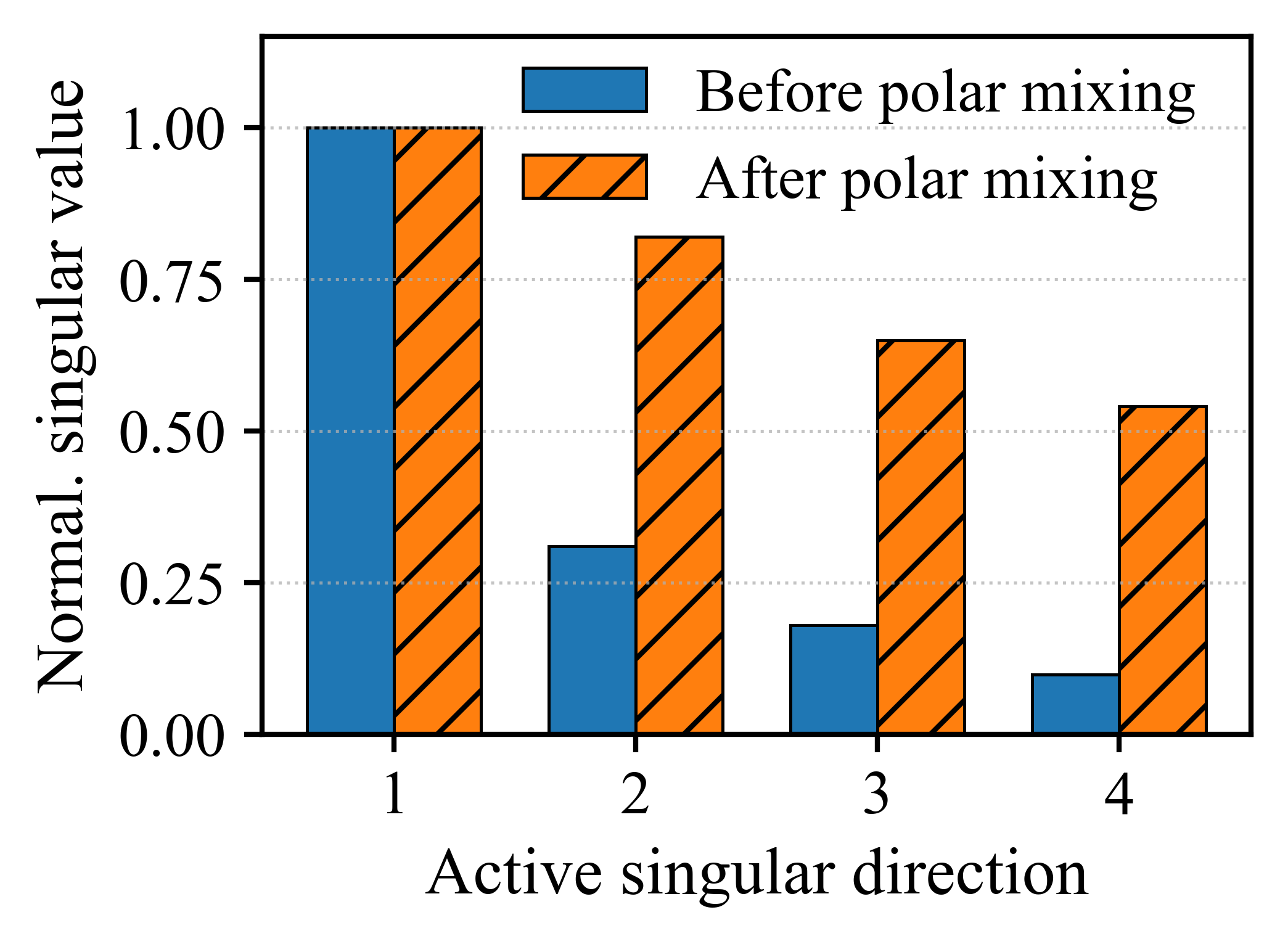}
        \caption{Active singular spectrum of a representative QNM-gate momentum block before and after Frobenius-preserving polar mixing. Before mixing, the spectrum is anisotropic, with a few dominant directions and several weak active singular directions. After mixing, the spectrum becomes flatter: weak active directions are lifted while the dominant structured directions are preserved. This visualizes the spectral-regularization mechanism established in Proposition~\ref{prop:floor} and Corollary~\ref{cor:cond}.}
        \label{fig:spectral_spectrum_appendix}
    \end{figure}

    Fig.~\ref{fig:spectral_spectrum_appendix} complements the main spectral summary by displaying the singular values themselves. The observed flattening is consistent with the theoretical role of polar mixing: it preserves the active singular subspaces of the structured momentum block while raising weak active directions through the additive spectral floor. This provides empirical evidence that QNM mitigates spectral collapse in circuit-semantic update blocks, consistent with the predicted mechanism.

\subsection{Supervised Classification Setup} 
\label{app:classification_details} 
    The supervised classification experiments use the Fisher Iris dataset as a controlled transfer diagnostic for the proposed optimizer. We consider two binary classification tasks. The first task separates setosa from the other two classes, while the second task separates versicolor from virginica after removing setosa examples. These two tasks provide different levels of difficulty: setosa is nearly linearly separable in standard feature space, whereas versicolor versus virginica is a harder binary discrimination task. Each train--test split is stratified so that the class proportions are preserved. Let  $x_i\in\mathbb R^4, \quad y_i\in\{0,1\}$
    denote an input feature vector and its binary label. To avoid data leakage, PCA is fit only on the training split. The same fitted PCA transformation is then applied to the corresponding test split. Specifically, if $\mathbf W_{\rm PCA}\in\mathbb R^{4\times2}$ denotes the two-dimensional PCA projection learned from the training data, then the reduced feature is $z_i = \mathbf W_{\rm PCA}^{\top} \left(x_i-\bar{x}_{\rm train}\right) \in\mathbb R^2$
    The two principal coordinates are then rescaled to quantum rotation angles. We denote the resulting encoded angle vector by $ \phi_i = \operatorname{scale}(z_i) \in\mathbb R^2$.
    The scaling map is fit on the training split and then reused on the test split, again to prevent test information from entering the training pipeline. The classifier is a two-qubit data-reuploading variational circuit. For an encoded input $\phi_i$, the circuit prepares a state 
    \begin{align} 
        \ket{\psi(\phi_i;\theta)} = \mathbf U(\phi_i;\theta)\ket{00}, 
    \end{align} 
    where $\mathbf U(\phi_i;\theta)$ alternates data-encoding rotations, trainable single-qubit rotations, and entangling gates. The trainable parameters are organized as a tensor $\boldsymbol{\Theta} \in \mathbb R^{2\times2\times3}$
    where the three axes correspond to variational layers, qubits, and trainable rotation axes. Equivalently, the flattened parameter vector $\theta$ has dimension $d = 2\cdot 2\cdot 3 = 12$.
    This tensor structure provides the circuit-semantic block geometry used by QNM-style spectral normalization. The model output is obtained from a measured Pauli expectation value. In the binary setting, we use a scalar logit of the form $h_\theta(\phi_i) = a\,\langle Z_0\rangle_{\phi_i,\theta}+b$
    where $a$ and $b$ are trainable or fitted readout parameters and $\langle Z_0\rangle_{\phi_i,\theta}$ denotes the expectation of the Pauli-$Z$ observable on the readout qubit. The predicted probability is $p_\theta(y_i=1\mid \phi_i) = \sigma\!\left(h_\theta(\phi_i)\right)$
    where $\sigma(\cdot)$ is the logistic sigmoid. Training minimizes the empirical binary cross-entropy loss 
    \begin{align} 
        C(\theta) = \frac{1}{N_{\rm train}} \sum_{i=1}^{N_{\rm train}} \left[ -y_i\log p_\theta(y_i=1\mid\phi_i) - (1-y_i)\log\!\left(1-p_\theta(y_i=1\mid\phi_i)\right) \right]. 
    \end{align} 
    For supervised classification, the QNM variants use an empirical supervised QGT/QFIM constructed from the circuit sensitivities over the training mini-batch. At iteration $t$, let $g_t=\nabla C(\theta_t)$ denote the loss gradient and let $\mathbf F_t^{\rm sup}$ denote the empirical supervised geometry matrix. The damped natural-gradient direction is computed from 
    \begin{align} 
        \mathbf G_t^{\rm sup} = \mathbf F_t^{\rm sup} + \lambda\mathbf I, \quad u_t = \left(\mathbf G_t^{\rm sup}\right)^{-1}g_t . 
    \end{align} 
    We specify the empirical supervised geometry used in the classification experiments. Let $\mathcal B_t$ denote the training mini-batch at iteration $t$. For each encoded input $\phi_i\in\mathcal B_t$, define the input-conditioned variational state 
    \begin{align} 
        \ket{\psi_i(\theta)} = \ket{\psi(\phi_i;\theta)} = \mathbf U(\phi_i;\theta)\ket{00}. 
    \end{align} 
    For circuit parameters $\theta_j$ and $\theta_k$, the sample-wise quantum geometric tensor is 
    \begin{align} 
        \mathbf Q_i(\theta)_{jk} = \left\langle \partial_j\psi_i(\theta) \middle| \partial_k\psi_i(\theta) \right\rangle - \left\langle \partial_j\psi_i(\theta) \middle| \psi_i(\theta) \right\rangle \left\langle \psi_i(\theta) \middle| \partial_k\psi_i(\theta) \right\rangle , 
    \end{align} 
    where $\partial_j\psi_i(\theta)$ denotes $\partial \ket{\psi_i(\theta)}/\partial\theta_j$. The empirical supervised QFIM used by QNM is the mini-batch average of the real part of this input-conditioned QGT: 
    \begin{align} 
        \mathbf F_t^{\rm sup} = \frac{1}{|\mathcal B_t|} \sum_{i\in\mathcal B_t} \operatorname{Re}\!\left(\mathbf Q_i(\theta_t)\right). 
    \end{align} 
    Equivalently, defining the projected tangent vector 
    \begin{align} 
        \ket{\xi_{i,j}(\theta)} = \left( \mathbf I - \ket{\psi_i(\theta)}\bra{\psi_i(\theta)} \right) \partial_j\ket{\psi_i(\theta)}, 
    \end{align} 
    we have 
    \begin{align} 
        \mathbf F_t^{\rm sup}[j,k] = \frac{1}{|\mathcal B_t|} \sum_{i\in\mathcal B_t} \operatorname{Re} \left\langle \xi_{i,j}(\theta_t) \middle| \xi_{i,k}(\theta_t) \right\rangle . 
    \end{align} 
    Thus, $\mathbf F_t^{\rm sup}$ measures the average local sensitivity of the data-dependent quantum states over the supervised training mini-batch. The diagonal variant uses $\mathbf F_{t,\rm diag}^{\rm sup} = \operatorname{diag}\!\left(\mathbf F_t^{\rm sup}\right)$
    while the damped geometry used for preconditioning is $\mathbf G_t^{\rm sup} = \mathbf F_t^{\rm sup} + \lambda\mathbf I $
    or, for Diag-QNM, $\mathbf G_{t,\rm diag}^{\rm sup} = \mathbf F_{t,\rm diag}^{\rm sup} + \lambda\mathbf I$
    The readout parameters $a$ and $b$, when optimized, are treated as ordinary Euclidean readout parameters and are not included in the circuit QGT block. The QNM spectral step is applied to the variational circuit parameter tensor $\boldsymbol{\Theta}$.
    Using this empirical supervised geometry, the same QNM terminology used in the main paper is applied to the supervised setting:: 
    \begin{itemize}[leftmargin=1.25em] 
        \item \textbf{Diag-QNM} uses a diagonal approximation of $\mathbf F_t^{\rm sup}$, followed by damped preconditioning and Muon-style spectral normalization on the trainable parameter tensor. 
        \item \textbf{$\DAQNM$} applies distance-aware radius control to the normalized supervised QNM direction, using the same radius mechanism as in the main algorithm. 
    \end{itemize} 
    This experiment is intentionally small-scale. Its purpose is not to claim state-of-the-art supervised quantum classification performance on Iris, but to test whether the optimizer components transfer from variational quantum optimization objectives to a supervised empirical-loss setting. The exact training results in the main paper evaluate the deterministic optimizer behavior, while the finite-shot variants in Appendix~\ref{app:finite_shot_results} test robustness when expectation values and gradients are estimated from sampled measurements.

\subsection{Finite-Shot Supervised Results}
\label{app:finite_shot_results}
    The main paper focuses on deterministic descent, perturbation stability, and stationarity of the normalized distance-aware update. Finite-shot supervised classification is therefore reported here as an empirical robustness study, rather than as a theoretically characterized probabilistic shot-noise guarantee. The goal is to test whether the proposed geometric updates remain useful when expectations, gradients, and empirical geometry estimates are computed from sampled measurements.

    Training is performed either with exact expectation values or with finite-shot sampled observables. After training, final test metrics are evaluated exactly. Thus, the reported accuracies isolate the effect of finite-shot sampling noise on the optimization trajectory, rather than confounding it with sampling noise at evaluation time. All methods are trained for $70$ optimization steps and evaluated over five independent seeds. We report mean test accuracy in percent together with the standard error across seeds.

    \begin{table*}[h]
    \centering
    \small
    \setlength{\tabcolsep}{3pt}
    \caption{Finite-shot training sweep on the two Iris binary classification tasks. Training uses either exact expectation values or finite-shot sampled observables, while final test metrics are evaluated exactly after $70$ optimization steps. Results are mean test accuracy in percent $\pm$ standard error over five seeds.}
    \label{tab:shot_sweep_iris_combined_appendix}
    \begin{tabular}{llccc}
    \toprule
        Task & Shots & Adam & Diag-QNM & \DAQNM\\
    \midrule
    \multirow{4}{*}{Setosa vs. rest}
        & Exact & $91.85\pm0.74$ & $\mathbf{93.33\pm1.28}$ & $\mathbf{93.33\pm1.28}$ \\
        & 32    & $43.70\pm11.64$ & $\mathbf{52.59\pm9.97}$ & $44.44\pm6.79$ \\
        & 128   & $51.11\pm5.88$ & $\mathbf{55.56\pm3.85}$ & $53.33\pm4.63$ \\
        & 512   & $54.81\pm6.06$ & $60.00\pm7.80$ & $\mathbf{68.15\pm4.51}$ \\
    \midrule
    \multirow{4}{*}{Versicolor vs. virginica}
        & Exact & $80.56\pm4.01$ & $83.33\pm3.85$ & $\mathbf{86.67\pm1.92}$ \\
        & 32    & $\mathbf{45.56\pm7.29}$ & $34.44\pm7.78$ & $41.11\pm8.01$ \\
        & 128   & $\mathbf{60.00\pm13.33}$ & $48.89\pm4.01$ & $52.22\pm4.84$ \\
        & 512   & $56.67\pm5.09$ & $48.89\pm10.94$ & $\mathbf{63.33\pm8.39}$ \\
    \bottomrule
    \end{tabular}
    \end{table*}

    Table~\ref{tab:shot_sweep_iris_combined_appendix} reports the finite-shot supervised classification sweep. The exact-training rows first show that QNM-style quantum-geometric updates remain competitive beyond variational energy optimization and weighted QAOA. On setosa versus rest, Diag-QNM and $\DAQNM$ both achieve $93.33\%$ test accuracy, compared with $91.85\%$ for Adam. On the harder versicolor-versus-virginica task, $\DAQNM$ obtains the best exact-training result, reaching $86.67\%$, compared with $80.56\%$ for Adam and $83.33\%$ for Diag-QNM. These results support the interpretation that QGT/QFIM-informed spectral updates can transfer to supervised empirical-loss objectives, although this Iris experiment should be viewed as a controlled diagnostic rather than a broad supervised-QML benchmark.

    The finite-shot rows reveal a more nuanced pattern. At low and intermediate shot budgets, richer quantum-geometric updates are not uniformly beneficial. On setosa versus rest, Diag-QNM is strongest at $32$ and $128$ shots, reaching $52.59\%$ and $55.56\%$, respectively. This suggests that a diagonal empirical geometry can provide useful stabilization when the measurement budget is limited, because it introduces quantum-geometric preconditioning while avoiding the higher estimation burden of richer geometry. On the harder versicolor-versus-virginica task, Adam remains the strongest method at $32$ and $128$ shots, reaching $45.56\%$ and $60.00\%$, respectively. This behavior is consistent with the fact that finite-shot noise affects not only the gradient estimate but also the empirical geometry used by QNM-style updates.

    At the largest finite-shot budget, $\DAQNM$ becomes the strongest method on both tasks. With $512$ shots, $\DAQNM$ reaches $68.15\%$ on setosa versus rest, compared with $54.81\%$ for Adam and $60.00\%$ for Diag-QNM. On versicolor versus virginica, $\DAQNM$ reaches $63.33\%$, compared with $56.67\%$ for Adam and $48.89\%$ for Diag-QNM. Thus, at the most informative finite-shot setting tested here, $\DAQNM$ improves over Adam by $13.34$ percentage points on setosa versus rest and by $6.66$ percentage points on versicolor versus virginica.

    Overall, the finite-shot experiment suggests that the benefit of distance-aware quantum-geometric spectral updates depends on the quality of the sampled gradient and geometry estimates. When the shot budget is too small, simpler diagonal geometry or even Adam can be more robust. As the shot budget increases, the empirical QGT/QFIM and gradient estimates become more informative, allowing the richer $\DAQNM$ update to exploit preconditioning, spectral normalization, and radius control more effectively. We therefore interpret Table~\ref{tab:shot_sweep_iris_combined_appendix} as finite-shot robustness evidence for the proposed optimizer, while leaving a full high-probability convergence analysis under shot noise to future work.

    

    \begin{figure}[h]
    \centering
    \begin{subfigure}[t]{0.45\columnwidth}
        \centering
        \includegraphics[width=\linewidth]
        {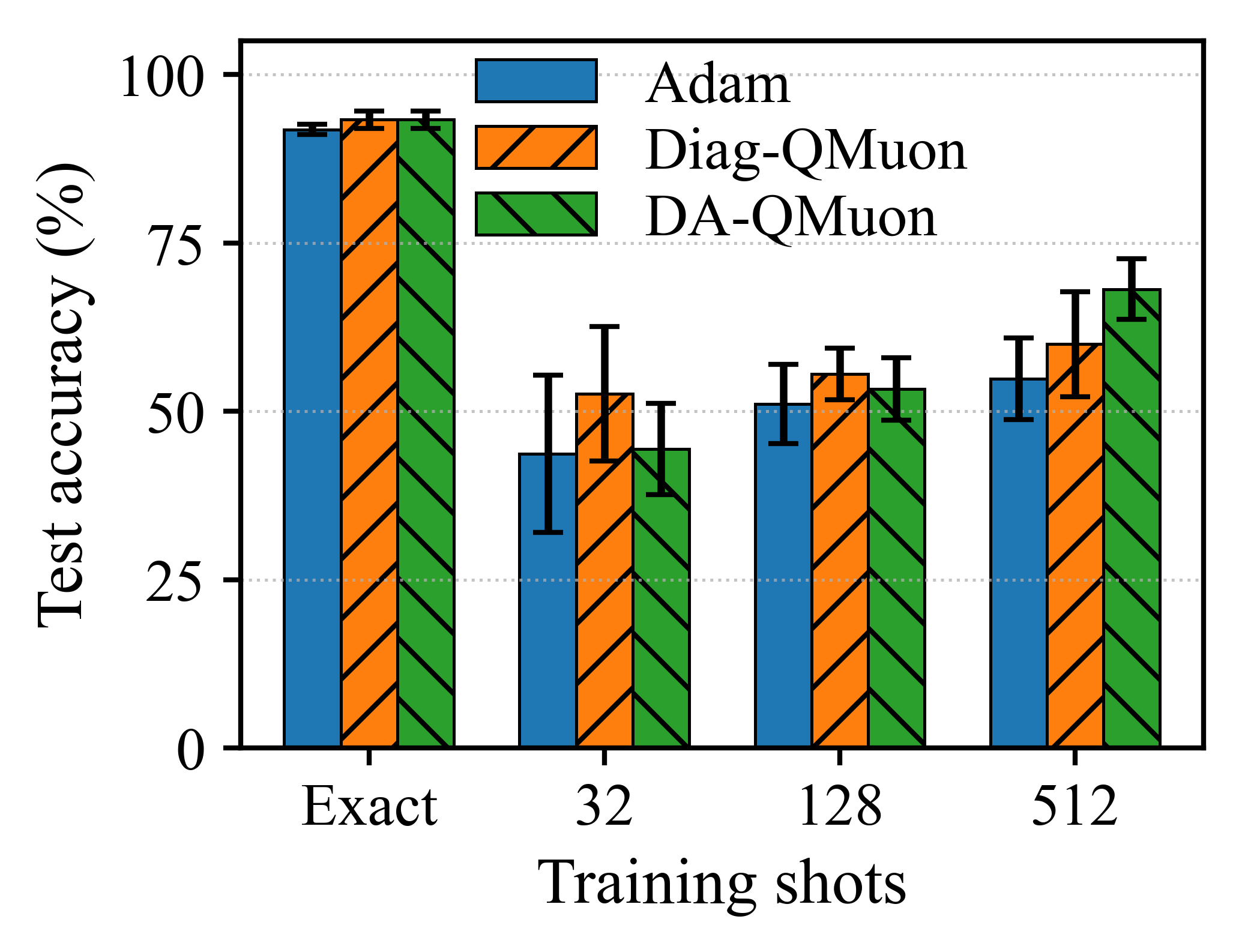}
        \caption{Setosa-versus-rest task.}
        \label{fig:finite_shot_setosa}
    \end{subfigure}
    \hfill
    \begin{subfigure}[t]{0.45\columnwidth}
        \centering
        \includegraphics[width=\linewidth]
        {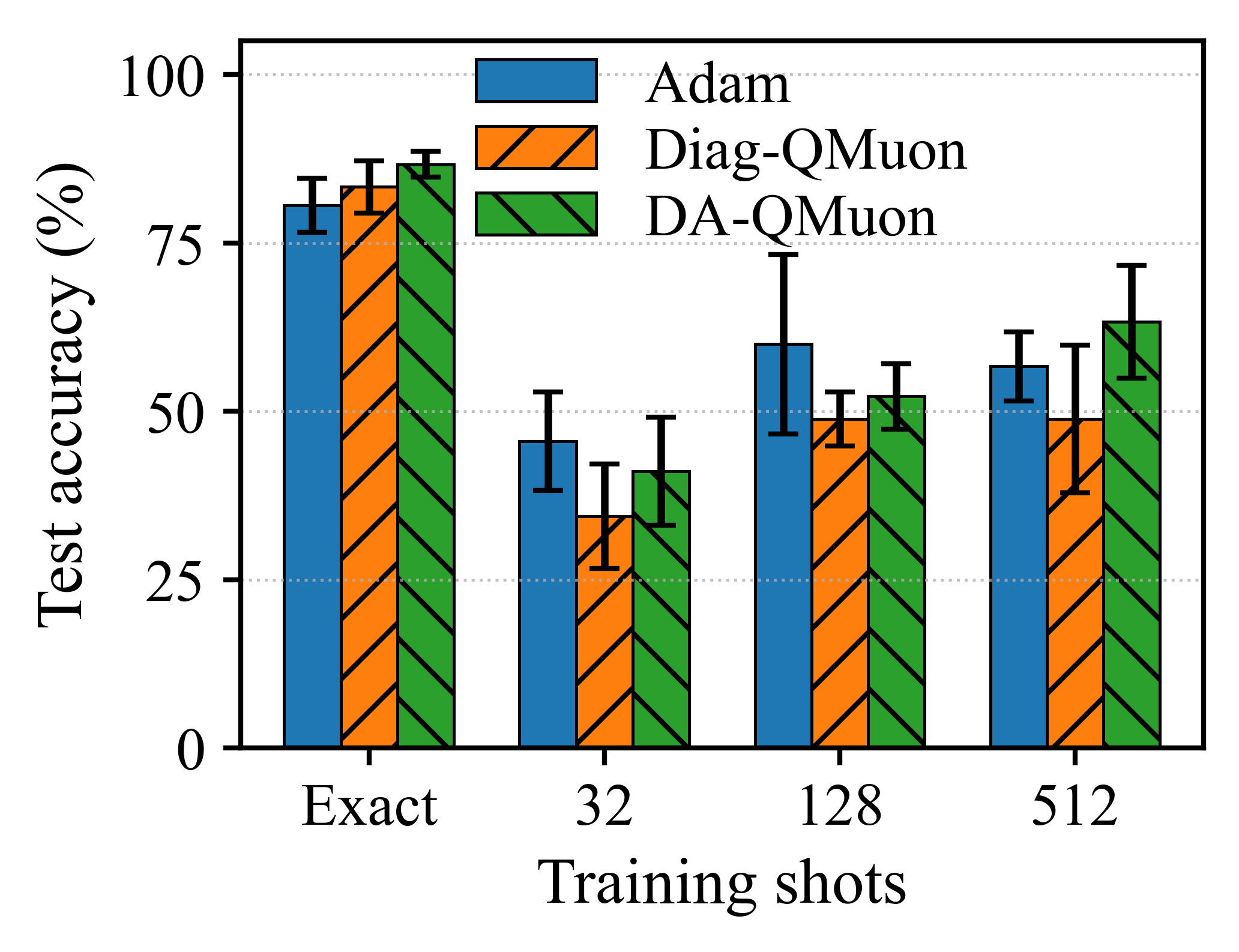}
        \caption{Versicolor-versus-virginica task.}
        \label{fig:finite_shot_versicolor}
    \end{subfigure}

    \caption{Finite-shot Iris classification after $70$ optimization
    steps. Training uses either exact expectation values or finite-shot
    sampled observables, while final test metrics are evaluated exactly.
    Error bars denote the standard error over five seeds.
    (a) Diag-QNM performs best at low and intermediate shot budgets,
    whereas $\DAQNM$ achieves the highest accuracy at $512$ shots.
    (b) Adam remains competitive at low and intermediate shot budgets,
    while $\DAQNM$ achieves the highest accuracy in the exact and
    $512$-shot settings.}
    \label{fig:finite_shot_iris}
\end{figure}
    
    Figs.~\ref{fig:finite_shot_setosa} and \ref{fig:finite_shot_versicolor} visualize the shot-budget dependence reported in Table~\ref{tab:shot_sweep_iris_combined_appendix}. The setosa task shows that diagonal empirical geometry can be useful when the measurement budget is limited: Diag-QNM performs best at $32$ and $128$ shots. In contrast, the richer distance-aware update becomes most effective at $512$ shots, where $\DAQNM$ achieves the highest accuracy. The harder versicolor-versus-virginica task shows a similar transition: Adam is competitive when the shot budget is small, but $\DAQNM$ becomes the strongest method once the finite-shot estimates are sufficiently informative. These patterns support the interpretation that QFIM preconditioning, spectral normalization, and distance-aware radius control are complementary, but that their combined benefit depends on the quality of the sampled gradient and geometry estimates.

\subsection{Implementation and Reproducibility} 
\label{app:reproducibility} 
    All exact-expectation experiments use statevector simulation. Finite-shot supervised experiments use sampled observables during training, while final test metrics are evaluated exactly after training. This protocol isolates optimization noise due to finite-shot estimation from sampling noise at evaluation time. The VQE ground-truth energies are computed by exact diagonalization, and the weighted QAOA optimum is computed by exhaustive enumeration for the $n=10$ weighted MaxCut instance. Unless otherwise stated, results are reported as mean $\pm$ standard error over the indicated seeds or splits. 
    
\paragraph{Implementation protocol.} 
    All optimizers are evaluated on matched task instances, matched random seeds, and the same initialization whenever applicable. For deterministic statevector experiments, observed differences therefore reflect optimizer behavior rather than measurement noise. We report best-achieved performance when the goal is to test whether an optimizer reaches a good basin, and final performance when the goal is to assess convergence stability. 
    
\paragraph{Task-specific implementation details.} 
    For the VQE experiment, we use a four-layer hardware-efficient ansatz for a transverse-field Ising Hamiltonian with periodic boundary conditions. The reported metric is the best energy error achieved during training, $\min_t \big(E(\theta_t)-E_\star\big)$
    where $E_\star$ is obtained by exact diagonalization. The VQE results are reported over seeds $0,1,2$, and the selected best QNM variant is QNM-layer for all three seeds. For the weighted QAOA experiment, the instance has $n=10, |E|=23, p=3$. The representative weighted QAOA run used matched initialization across optimizers. The DA-QNM QAOA implementation used CPU-based double-precision simulation, seed $0$, $180$ optimization steps, learning rate $0.04$, Adam learning rate $0.05$, momentum coefficient $\beta=0.9$, QFIM damping $\lambda=0.05$, QFIM refresh period $3$, maximum update norm $0.05$, and polar-mixing coefficient $\alpha=0.3$. For supervised quantum classification, we use the Fisher Iris dataset with two binary tasks: setosa versus rest and versicolor versus virginica. Exact training results are reported over stratified train--test splits. The finite-shot supervised experiments train each method for $70$ optimization steps and evaluate over five independent seeds. Training uses either exact expectation values or finite-shot sampled observables, while final metrics are evaluated exactly. 


\end{document}